%% file: paper_v3.tex
\documentclass[10pt]{elsarticle}
 
\usepackage{hyperref}

\usepackage{calrsfs,amsmath,amssymb,graphicx,array,accents,figlatex,epsfig}
\usepackage{setspace,wrapfig,colortbl,marvosym, bbm, stmaryrd,centernot, color,upgreek}
\usepackage{amsfonts}
\usepackage{epstopdf}
\usepackage{booktabs}
\usepackage{adjustbox}
\usepackage{multirow}
\usepackage{mathrsfs}
\usepackage{algorithm}
\usepackage{algpseudocode}
\usepackage{amsmath}
\usepackage{mathptmx}
\usepackage{newtxmath}
\usepackage{algorithm,algcompatible,amsmath}
\usepackage{graphicx}
\usepackage{dcolumn}
\usepackage{caption}
\usepackage{amsmath}
\usepackage{graphicx}
\usepackage{titlesec}

\usepackage[T1]{fontenc}
\usepackage{ae,aecompl}

\usepackage{fancyhdr}

\usepackage{etoolbox} 
\usepackage{natbib} 
\usepackage{blindtext}  
\usepackage{enumitem}

\ifpdf
  \DeclareGraphicsExtensions{.eps,.pdf,.png,.jpg}
\else
  \DeclareGraphicsExtensions{.eps}
\fi

\usepackage{enumitem}
\setlist[enumerate]{leftmargin=.5in}
\setlist[itemize]{leftmargin=.5in}

\newtheorem{rem}{Remark}

\input{macros}

\begin{document}

\begin{frontmatter}

\title{Network scaling and scale-driven loss balancing \\ for intelligent poroelastography} 

\author{Yang Xu$^1$}
\author{Fatemeh Pourahmadian$^{1,2}$\corref{cor1}}

\address{\vspace*{1.25mm}$^1$ Department of Civil, Environmental \& Architectural Engineering, University of Colorado Boulder, USA} 
\address{\vspace*{-1mm}$^2$ Department of Applied Mathematics, University of Colorado Boulder, USA}
\cortext[cor1]{Corresponding author: tel. 303-492-2027, email {\tt fatemeh.pourahmadian@colorado.edu}}

\begin{abstract}
A deep learning framework is developed for multiscale characterization of poroelastic media from full waveform data which is known as poroelastography. Special attention is paid to heterogeneous environments whose multiphase properties may drastically change across several scales. Described in space-frequency, the data takes the form of focal solid displacement and pore pressure fields in various neighborhoods furnished either by reconstruction from remote data or direct measurements depending on the application. The objective is to simultaneously recover the six hydromechanical properties germane to Biot equations and their spatial distribution in a robust and efficient manner. Two major challenges impede direct application of existing state-of-the-art techniques for this purpose: (i) the sought-for properties belong to vastly different and potentially uncertain scales, and~(ii) the loss function is multi-objective and multi-scale (both in terms of its individual components and the total loss). To help bridge the gap, we propose the idea of \emph{network scaling} where the neural property maps are constructed by unit shape functions composed into a scaling layer. In this model, the unknown network parameters (weights and biases) remain of O(1) during training. This forms the basis for explicit scaling of the loss components and their derivatives with respect to the network parameters. Thereby, we propose the physics-based \emph{dynamic scaling} approach for adaptive loss balancing. The idea is first presented in a generic form for multi-physics and multi-scale PDE systems, and then applied through a set of numerical experiments to poroelastography. The results are presented along with reconstructions by way of gradient normalization (GradNorm) and Softmax adaptive weights (SoftAdapt) for loss balancing. A comparative analysis of the methods and corresponding results is provided. The case of multi-scale reconstructions from noisy data is also numerically investigated.   

\end{abstract}

\end{frontmatter}

\section{Introduction}

Multiphasic processes are at the heart of many critical applications including renewable energy~\cite{tang2024,boye2023}, carbon capture~\cite{yeri2024,anth2022}, and medicine~\cite{sowi2021,tan2016}. For instance, engineered injection of fluids in the subsurface, known as the stimulation process, has remarkably enhanced sustainable energy mining from geothermal reservoirs~\cite{Sare2009,Shev2014} offering an important complement to other renewables with the potential of generating over 100 GW$_e$ of cost-competitive capacity by 2050~\cite{renner2006,natio2006}. Realizing this potential relies on smart stimulation schemes that require {\emph{real-time feedback on the subsurface hydromechanical evolution}}~\cite{de2024}. The latter is particularly crucial in enhanced geothermal systems owing to their continuous demand for stimulation and recharge for long-term production~\cite{renner2006}. In magnetic resonance elastography~\citep{muth1995}, a medical imaging technique for in vivo mechanical characterization of soft tissues, it is well known that single-phase elastodynamic models fail to capture the true behavior of live tissues with high water content. This has reportedly led to unstable and unverifiable reconstructions of brain (and other organs)~\cite{perr2010,mcga2015,tan2017}. Recent developments highlight the utility and importance of Biot's poroelastodynamics theory~\cite{biot1956(1),biot1956,biot1962} in capturing the fluid-solid interactions in live tissues and is shown to result in more accurate and stable reconstructions~\cite{meyer2024,gala2023,hoss2020}. {\emph{In vivo hydrodynamical characterization}} of biological materials could significantly aid differential diagnosis and progress monitoring of neurological disorders such as hydrocephalus and brain tumors~\cite{patt2014}. 

Early poroelastograms referred to time sequences of effective Poisson's ratio calculated from the evolving ratio of radial to axial strains in poroelastic materials during quasi-static stress relaxation~\cite{kono2001,berr2006}. This information could then be used to quantify the specimen's permeability. However, this approach involved major simplifying assumptions on the physics of fluid motion and solid deformation in the sample. Recent advances in magnetic resonance motion sensing has enabled a more comprehensive characterization of biphasic materials from volumetric time-harmonic displacement data (mainly in the low-frequency regime) where nonlinear minimization is employed to reconstruct piece by piece the distribution of (a subset of) Biot parameters along with the interstitial fluid pressure fields~\cite{perr2010,patt2014,meyer2024}. The remaining challenges in this vein include tardy reconstructions, and sensitivity of the results to the motion frequency, noise in data, and other (poroelastic) properties of the specimen which are not directly identified but estimated a priori using other methods. To help bridge the gap, this study aims to carefully extend our recent work on ML-based elastography~\cite{xu2023,pour2018} to poroelastodynamics in order to enable simultaneous reconstruction of all Biot parameters from full-field displacement and pore pressure data in a target region of the subsurface. Special attention is paid to challenges related to multi-scale and multi-physics full waveform inversion as well as reconstruction from noisy data. It should be mentioned that this study is particularly related and may contribute to sequential remote sensing~\cite{pour2017} where the subsurface is first (geometrically) imaged by way of inverse scattering solutions~\cite{pour2022,liu2023} where subsurface regions of interest e.g., process zones in geo-energy systems or anomalous parts in soft tissues are identified for targeted hydromechanical characterization. Next, local (pore pressure and elastic displacement) fields will be recovered in the identified zones of interest from remote measurements using recent developments in auto-focusing~\cite{wape2023}. This may be accomplished by Marchenko-type integral equations where measurements on the surface are deployed to generate focal fields at arbitrary points in a heterogeneous subsurface of unknown properties~\cite{wape2021}. Such focal fields yield components of the relevant Green's function as in~\cite{wape2023} (or its recent variants~\cite{boul2022,gin2021}) where the source location coincides with the targeted focus point. On repeating the algorithm for a number of focal points, one may generate sufficient instances of local wavefields in a region of interest for multi-phase characterization. It should be noted that in some applications such as convection MRI~\cite{sala2022,walk2018}, direct measurement of full poroelastic waveforms in the subsurface may also be an option. Given the above, the primary objective of this investigation is the piece-wise reconstruction of Biot properties from full poroelastic waveforms in a neighborhood of focal points.          

The reconstruction entails multitask learning~\cite{Caru1998,Bakk2003} where the proposed multiscale neural network, modeling poroelastic properties of the subsurface, minimizes the cumulative residual of a system of partial differential equations (PDEs). More specifically, the loss function takes the form $\mathcal{L}=\frac{1}{N}\sum_{i=1}^{N} \norms{\nxs w_i \ell_{i} \nxs}^2 $ wherein $\ell_{i}$ denotes a single-task objective with the associated weight $w_i$, and the summations is over $N$ PDEs. In multi-phase systems, $\ell_{i}$s describe physical principles that involve spatial and temporal scales straddling multiple orders of magnitude. Thus, careful tuning of the weights is paramount for a robust and successful reconstruction. In deep learning community, loss balancing for multitask learning goes beyond the limits of physics and has been the subject of mounting interest in a wide range of applications. In computer vision, multitask learning is leveraged to improve generalization capability of UberNet~\cite{Kokk2016} and Mask R-CNN~\cite{He2017} through shared representations for instance segmentation and object detection. Traditional methods of loss balancing typically use static and/or manually-tuned weights. For example, sparse autoencoders employ a small, constant weight, known as the sparsity parameter, for the regularization term which is usually found by trial and error that can be suboptimal and labor-intensive. Recent developments, however, have focused on optimizing the learning process by adaptively balancing the loss contributions of different tasks i.e., $w_i = w_i(t)$ where $t$ indicates the training step. Kendall et al.~\cite{Kend2017} introduced a method that calculates the weights based on the homoscedastic uncertainty in each task. This approach has been shown to outperform static weight baselines and improve overall task performance. Building on this idea, GradNorm~\cite{chen2018} was recently developed as a gradient normalization algorithm that balances training by adjusting gradient magnitudes dynamically. GradNorm aims to equalize the training rates of different tasks, ensuring that no single task dominates the learning process. This method has been shown to improve accuracy and reduce overfitting across tasks for various network architectures and datasets. Another approach is the so-called SoftAdapt~\cite{heyd2019} which uses a family of Softmax-inspired methods to adaptively update the weights based on the live statistics of each subloss. This technique makes use of the initial (or previous) iterations in each epoch to create a preconditioner matrix that normalizes the partial gradients of the loss function in the parameter space. SoftAdapt can be easily integrated into existing architectures, providing a flexible and efficient way to optimize multitask learning. GradNorm was recently adopted by~\cite{amin2023} for multiphysics characterization of the subsurface using physics-informed neural networks and reportedly resulted in unstable reconstructions. This inspired a new training logic based on splitting the PDE system into parts dominated by distinct physics/scales. The latter will work in many situations but may pose challenges in highly-heterogeneous systems where the nature of coupling between various physical phenomena may rapidly change in a small neighborhood. In addition, proper decomposition of a PDE system may be particularly complicated in dynamic problems with highly oscillatory data where training PINNs proved to be relatively slow and challenging even in systems governed by single-scale PDEs~\cite{xu2023}. In general, in multiscale and multiphysics systems, disparity in gradients exists and some features associated to small time and/or space scales i.e., high frequencies and wave numbers may be less observable and more impacted by noise compared to others. In light of this and given the errors that may be involved in network-estimated derivatives of data that appear in the loss components, especially in initial epochs, it seems that loss balancing based on normalized gradients may lead to amplifying noise or suppressing the signal that may explain the observed instabilities in recent works, and thus, may need to be revisited and adapted for such applications.

In our effort to extend ML-based elastography~\cite{xu2023} to poroelastography, we encountered two major impediments:~(1) the sought-for material properties, i.e., the outputs of neural network, belong to several different scales owing to their distinct physical nature; the scaling for some of the parameters may be known a priori but for others such as the permeability coefficient or porosity may be uncertain or vary on a set of scales, and~(2) various equations in the governing PDE system, i.e., the loss components, describe the balance and coupling between physical phenomena at largely different scales so that the normalized equations remain multiscale, both individually and as a system, such that for instance the leading order in one equation may be $O(1)$, while in the other is $O(10^{-3})$. The first challenge remains true even after standard non-dimensionalization of the governing equations by introducing unit reference scales for relevant physical quantities. For instance, after normalization, the shear modulus in a specimen may be of $O(1)$ while the permeability coefficient is of $O(10^{-5})$. One solution is hardcoding via a change of variable so that all the PDE parameters are re-scaled to $O(1)$. This has been successfully implemented for instance in~\cite{amin2023}. A caveat of this approach is that (a) it is not applicable in heterogeneous environments where some of the unknowns such as porosity or permeability could span several scales, and (b) the exact scaling of each unknown quantity should be known a priori. 
 
This work presents a few ideas that could potentially address the above challenges. This paper is organized as the following. Section~\ref{PPA} introduces the proposed network scaling and scale-driven loss balancing in a generic framework for multi-scale PDE systems. Section~\ref{MP} describes the problem statement for poroelastography and adapts the proposed approach for this application. In addition, the state-of-the-art GradNorm and SoftAdapt techniques for loss balancing are briefly introduced. This section also includes an analysis of the proposed dynamic scaling and its relation to GradNorm and SoftAdapt. Section~\ref{ER} provides an account of numerical experiments where the reconstruction results from the three approaches are compared and the case of reconstruction from noisy data is investigated. This is then followed by a detailed discussion of the observations and conclusions.

\section{Proposed approach}\label{PPA}

\emph{Network scaling} embeds a generic change of variable in the architecture of neural networks that map the unknown physical properties of a system in space. In this approach a scaling layer is added right before the output while other layers are normalized i.e.,~scaled to $O(1)$. For instance, let us consider a multilayer perceptron (MLP) as a property map where the input consists of a grid of neighborhoods in space $\boldsymbol{\xi}_i$, $i = 1,2, \ldots, N_\xi$, while the output is the associated material properties (or PDE parameters) in each neighborhood $\boldsymbol{\vartheta}_i = (\vartheta_1, \vartheta_2, \ldots, \vartheta_{N_{\boldsymbol{\vartheta}}} )_i$ where ${N_{\boldsymbol{\vartheta}}}$ is the number of PDE parameters. In this setting, every MLP layer except the last one is defined by the standard map $\Upsilon_\ell\! : \boldsymbol{x}^{\ell-1} \to \boldsymbol{x}^{\ell}$,
\begin{equation}
\label{eq:layer}
\boldsymbol{x}^{\ell} ~=~ \Upsilon_\ell\big{(}\boldsymbol{x}^{\ell-1}\big{)} ~\colon\!\!\!\!\!=~ \sigma\big{(}\boldsymbol{W}^{\ell} \boldsymbol{x}^{\ell-1} +\, \boldsymbol{b}^{\ell}\big{)}, \qquad \ell = 1,2, \ldots, N_\ell - 1,
\end{equation}
where $\boldsymbol{W}^\ell $ and $\boldsymbol{b}^\ell$ respectively designate the $\ell^\text{th}$ layer's weight and bias and $\sigma =$ tanh, ReLU is the activation function. The network is constructed by coherent composition of $\Upsilon_\ell$ for $\ell = 1,2, \ldots, N_\ell - 1$, wherein $N_\ell$ denotes the number of layers. In the last layer, the scaling is applied as the following 
\begin{equation}
\label{eq:last_layer}
\big{(}\boldsymbol{\vartheta}_1, \boldsymbol{\vartheta}_2, \ldots, \boldsymbol{\vartheta}_{\nxs N_\xi}\!\big{)}~=~ \text{diag}\big{[} (s_1, s_2, \ldots, s_{N_{\boldsymbol{\vartheta}}} )_1, \ldots, (s_1, s_2, \ldots, s_{N_{\boldsymbol{\vartheta}}} )_{\nxs N_\xi}\big{]} \boldsymbol{x}^{N_\ell - 1},
\end{equation} 
wherein $(s_i)_j$ denotes the scale of $\vartheta_i$, $i = 1,\ldots, {N_{\boldsymbol{\vartheta}}}$, in the $j^\text{th}$ neighborhood, $j = 1,\ldots, N_\xi$, such that $(\vartheta_i)_j = O\big{(}(s_i)_j\big{)}$. In most physical systems, the majority of system's properties vary in space but remain of the same scale i.e., $(s_i)_j = s_i, \forall j, i \in I \subset \lbrace 1,\ldots, {N_{\boldsymbol{\vartheta}}} \rbrace$ except for a small subset of quantities in which case the scaling factors may be identified by a small set $\mathbb{S}_i$ of relevant scales (indicating the likely range of scales that the associated property may assume), thereby the correct scaling factor in each neighborhood will be selected during the learning process such that $(s_i)_j \in \mathbb{S}_i, \forall j, i \in \lbrace 1,\ldots, {N_{\boldsymbol{\vartheta}}} \rbrace \nxs\setminus\nxs I$. For example, the scaling for permeability in rocks can assume values from the set $\lbrace 10^{-5}, 10^{-6}, 10^{-7} \rbrace$ and the exact factors in various neighborhoods will be picked from this set during the optimization process. Note that in the proposed architecture, the network parameters i.e., weights and biases remain of $O(1)$ which aids their fast and robust identification, while the output is properly scaled by the last layer. Note that the scaling factors in this layer are either exactly determined by the nominal values or preliminary experiments, or belong to a small set of scales specified based on the expected range of variation for the associated physical quantities in the domain.   

Given the system's response $\boldsymbol{v}$, over a set of neighborhoods, the objective is to learn the unknown parameters of the property map ($\lbrace \boldsymbol{W}^\ell, \boldsymbol{b}^\ell \rbrace$, $\ell = 1,2, \ldots, N_\ell - 1$, and $(s_i)_j \in \mathbb{S}_i, \forall j, i \in \lbrace 1,\ldots, {N_{\boldsymbol{\vartheta}}} \rbrace \nxs\setminus\nxs I$) by minimizing the residual of the system's governing equations $\mathcal{L}=\frac{1}{N}\sum_{i=1}^{N} \norms{\nxs w_i \ell_{i} \nxs}^2 $. It should be mentioned that here we consider data inversion in the frequency-domain i.e.,~$\boldsymbol{v} = \boldsymbol{v}(\bx,\omega)$ and since multiscale physical quantities are frequency-dependent, we will focus on reconstructions at fixed frequencies. In presence of multi-frequency data, one may conduct the inversion process at each frequency separately and uncover the PDE parameters as a function of frequency (or their evolution in time by an inverse Fourier transform). We assume that each equation $\ell_i$ in the PDE system can be expressed as a summation of separable operators in terms of data $\boldsymbol{v}$ and the unknown parameters $\boldsymbol{\vartheta}$ such that    
\beq\lb{Fact}
\ell_{i} = \sum_{j=1}^{N_j} f_{ij}(\boldsymbol{\vartheta},\omega) D_{ij}(\boldsymbol{v}), \quad D_{ij}(\boldsymbol{v}) = \sum_{k=1}^{N_k} \frac{\partial^{\alpha_1^k+\alpha_2^k+\alpha_3^k}}{\partial x_1^{\alpha_1^k} \partial x_2^{\alpha_2^k}  \partial x_3^{\alpha_3^k} } h_{ij}^k[\boldsymbol{v}](\bx), \quad \boldsymbol{\alpha}^k = (\alpha_1^k, \alpha_2^k, \alpha_3^k),
\eeq         
wherein $f_{ij} = f_{ij}(\boldsymbol{\vartheta},\omega)$ and $h_{ij} = h_{ij}^k(\boldsymbol{v})$ for $i =1,\ldots, N$, $j = 1,\ldots, N_j$ and $k = 1,\ldots, N_k$, are linear or nonlinear functions of their arguments; $D_{ij}$ denotes a differential operator, and the multi-index notation $\boldsymbol{\alpha}^k \in \mathbb{N}^3$ is used for partial differentiation with respect to $\bx = (x_1, x_2, x_3)$~\cite{schm2024}.     

In this setting, for \emph{loss balancing}, by assuming that $f_{ij}$ is a Lipchitz function of network parameters (see Remark~\ref{rm1}), we propose to scale each loss component $\ell_i$ by a dynamic weight $w_i$ such that the average scale in the normalized equation $w_i \ell_{i}$ is $O(1)$. For this purpose, at every epoch, the scale of every loss term in~\eqref{Fact} is determined as the following
\beq\lb{Scl}
e_{ij}~=~\text{round}\big{(}\text{log}_{10}[f_{ij}(\boldsymbol{\vartheta},\omega)]\big{)} \,+\, \text{round}\big{(}\text{log}_{10}[\langle | D_{ij}(\boldsymbol{v}) | \rangle_{\bxi}]\big{)}, \quad i~=~1,\ldots, N, \,\, j~=~1,\ldots, N_j,
\eeq         
where round() maps its argument to the nearest integer and $\langle | \cdot | \rangle_{\bxi}$ denotes the mean of absolute value of its argument over the support $\bxi$. Note that the scale of $f_{ij}\big{(}\boldsymbol{\vartheta},\omega)$ is \emph{in part} dictated by the last layer in the neural property map and can also be determined by the network's estimates for the PDE parameters $\boldsymbol{\vartheta}$. Here, $\omega$ is the frequency of probing waves and a known (input) parameter. Moreover, the scale of $D_{ij}(\boldsymbol{v})$ is directly computed from data using spectral methods for signal processing and differentiation~\cite{xu2023,pour2018}. Given the above, the average scale and affiliated weight for each loss component is specified as follows, 
\begin{equation}
\label{eq:dynamicweights}
w_i ~=~ 10^{\exs - \beta_{i}}, \quad \beta_{i} ~=\,  \displaystyle \frac{1}{N_j} \sum_{j=1}^{N_j}e_{ij}, \quad i ~=~1,\ldots, N.
\end{equation}
Here we assumed that $10$ is an appropriate basis for separating the scales. One may replace $10$ by $10^{\eta}$ and gauge the basis as needed in their application of interest. The pseudocode for the proposed approach is provided in Algorithm~1.

\begin{algorithm}[tp!]
\label{AL1}
\caption{Pseudocode for the method of \emph{dynamic scaling} for loss balancing.}
  \begin{algorithmic}
    \REQUIRE (1)~network-predicted scaling for unknown physical parameters $\boldsymbol{\vartheta}$,~(2)~loss components $\ell_i$ recast according to~\eqref{Fact}, and (3) relevant field derivatives $D_{ij}(\boldsymbol{v})$ that appear in the loss

    \vspace*{1 mm}
    \FOR{$t = 0$ \textbf{to} epoch}
    \vspace*{1 mm}
     \IF{$f_{ij}$ is Lipchitz continuous $\forall i,j$ with respect to the network parameters}
      \vspace*{2 mm}
      \STATE Compute $\text{round}\big{(}\text{log}_{10}[f_{ij}(\boldsymbol{\vartheta},\omega)]\big{)}$ for $i = 1,\ldots, N$ and $j = 1,\ldots, N_j$ based on the network outputs
    \vspace*{2 mm}  
      \STATE Compute $\forall i,j$, $\text{round}\big{(}\text{log}_{10}[\langle | D_{ij}(\boldsymbol{v}) | \rangle_{\bxi}]\big{)}$ according to the loss 
      \vspace*{2 mm} 
      \STATE Compute the average scale of each loss component \vspace*{-2mm}$$ \beta_{i} ~=\, \displaystyle \frac{1}{N_j} \sum_{j=1}^{N_j} \Big{(} \text{round}\big{(}\text{log}_{10}[f_{ij}(\boldsymbol{\vartheta},\omega)]\big{)} \,+\, \text{round}\big{(}\text{log}_{10}[\langle | D_{ij}(\boldsymbol{v}) | \rangle_{\bxi}]\big{)} \Big{)}$$. \vspace*{-5mm}
    \ELSE
       \vspace*{2 mm}
       \STATE Compute $\text{round}\big{(}\text{log}_{10}[\langle | D_{ij}(\boldsymbol{v}) | \rangle_{\bxi}]\big{)}$ for $i = 1,\ldots, N$ and $j = 1,\ldots, N_j$ according to the loss
    \vspace*{2 mm}  
        \STATE Compute $\forall i,j, n$, $\text{{round}}\left(\text{{log}}_{10}\!\left[\frac{\partial f_{ij}}{\partial \vartheta_n}({\vartheta_n},\omega)\right]\right)$, $n = 1,\ldots, N_{\boldsymbol{\vartheta}}$, based on the network outputs
        \vspace*{2 mm}
        \STATE Compute $\forall n$, $\text{{round}}\big{(}\text{{log}}_{10}[\vartheta_n]\big{)}$, based on the network outputs
      \vspace*{4 mm} 
      \STATE Compute the average scale of each loss component's derivative with respect to a network parameter \vspace*{-1mm}$$ \beta_{i} ~=\, \displaystyle \frac{1}{N_j} \! \sum_{j=1}^{N_j} \left( \text{{round}}\big{(}\text{{log}}_{10}[\langle | D_{ij}(\boldsymbol{v}) | \rangle_{\bxi}]\big{)} \,+\, \frac{1}{N_{\boldsymbol{\vartheta}}} \sum_{n=1}^{N_{\boldsymbol{\vartheta}}} \left( \text{{round}}\left(\text{{log}}_{10}\!\left[\frac{\partial f_{ij}}{\partial \vartheta_n}({\vartheta_n},\omega)\right]\right) \,+\, \text{{round}}\big{(}\text{{log}}_{10}[\vartheta_n]\big{)} \right) \right)$$. \vspace*{-3mm}
   	\ENDIF 
	\vspace*{1 mm}  
      \STATE Update weights $w_i = \displaystyle {10^{- \beta_{i}}}$
      \STATE Update weighted loss $\mathcal{L} = \displaystyle \frac{1}{N}\sum_{i=1}^{N}\norms{\nxs w_i \ell_{i} \nxs}^2$
      \STATE Optimize $\mathcal{L}$
	 \vspace*{1 mm}  

    \ENDFOR 
  \end{algorithmic} 
\end{algorithm}
  
\begin{rem}\label{rm1}
Let $\text{\emph{w}}$ be a network parameter, observe that if $f_{ij}$ is Lipchitz continuous with respect to $\text{\emph{w}}$ such that the Lipchitz constant is of $O\left(f_{ij}\right)$, thanks to the proposed architecture for property maps and their scaling, then the proposed weights $w_i$ in~\eqref{eq:dynamicweights} automatically normalize the gradient of weighted loss components with respect to the network parameter $\text{\emph{w}}$. This will be demonstrated in Section~\ref{MP} within the context of poroelastography. In other applications where $f_{ij}$ is not Lipchitz continuous, one may alternatively use the following weights for gradient normalization based on scaling
\begin{equation}
\label{eq:dynamicweights2}
w'_i = \displaystyle {10^{- \beta'_{i}}}, \quad \beta'_{i} ~=\, \displaystyle \frac{1}{N_j} \! \sum_{j=1}^{N_j} \left( \text{\emph{round}}\big{(}\text{\emph{log}}_{10}[\langle | D_{ij}(\boldsymbol{v}) | \rangle_{\bxi}]\big{)} \,+\, \frac{1}{N_{\boldsymbol{\vartheta}}} \sum_{n=1}^{N_{\boldsymbol{\vartheta}}} \left( \text{\emph{round}}\left(\text{\emph{log}}_{10}\!\left[\frac{\partial f_{ij}}{\partial \vartheta_n}({\vartheta_n},\omega)\right]\right) \,+\, \text{\emph{round}}\big{(}\text{\emph{log}}_{10}[\vartheta_n]\big{)} \right) \right),
\end{equation}
where the scale of $\vartheta_n$ is governed by the last layer in the property network. Note that given the architecture of the latter, $\frac{\partial \vartheta_n}{\partial \text{\emph{w}}} = O(\vartheta_n)$ for $n = 1,\ldots, {N_{\boldsymbol{\vartheta}}}$. 
\end{rem}

\section{Model problem}\label{MP}

This section implements the proposed approach for poroelastic characterization of rocks using elastic waveform and acoustic pore pressure data. In what follows, we begin with the problem statement and highlight its multiscale nature that introduced many challenges in our early attempts for data inversion using the available techniques. This is followed by proper normalization and construction of the proposed neural architecture with network scaling. For loss balancing, we provide a comparative analysis between the proposed dynamic scaling and the current state-of-art methods, namely: the GradNorm and SoftAdapt, for multi-task learning.

\vspace*{-1mm}
\subsection{Problem Statement}\label{ps}

\begin{figure}[!bp]
\begin{centering}
\includegraphics[width=0.65\columnwidth]{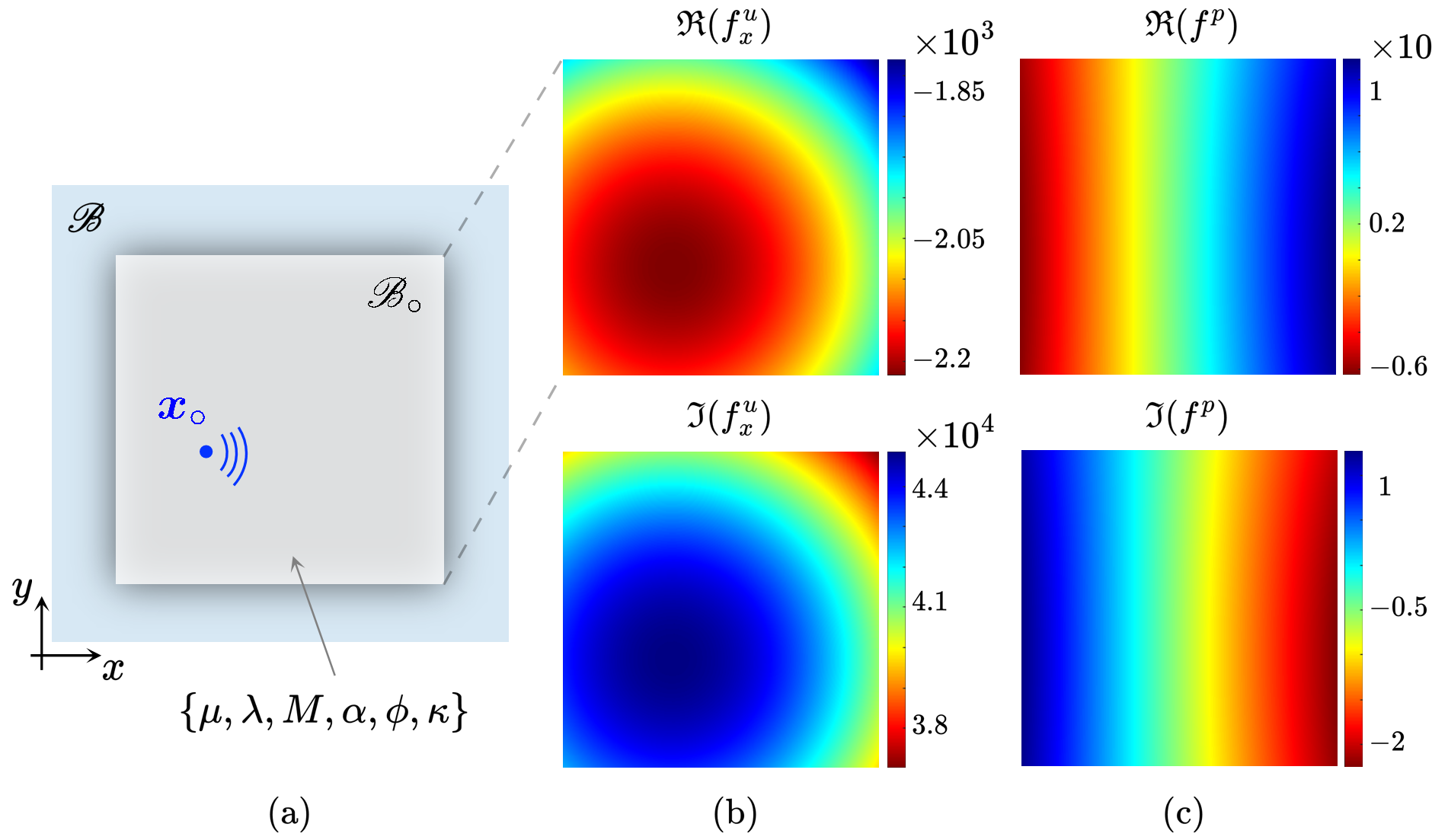}\vspace*{-2.5mm}
\par\end{centering}
\caption{Synthetic experiments simulating wave motion in the poroelastic domain $\mathcal{B}$:~(a) the model is harmonically excited at frequency $\omega$ by a fluid body force $\delta(\bx-\bx_\circ)$ and the response is computed in a neighborhood of the source point $\bx_\circ$ i.e., in the square $\mathcal{B}_\circ$ of side $5 \ell_r$,~(b) the $x$ component of effective body force $\boldsymbol{f}^{u}$ in the first of Biot equations in~\eqref{eq:Biot}, and~(c) the fluid source term $f^p$ in the generalized Darcy's law i.e., the second of~\eqref{eq:Biot}. Here, $\mathfrak{R}$ and $\mathfrak{I}$ respectively indicate the real and imaginary parts of a complex-valued quantity.}
\label{problem_statement1} \vspace*{-4mm}
\end{figure}

Consider a poroelastic domain $\mathscr{B}$ characterized by the drained Lam\'{e} parameters $\lambda$ and $\mu$, Biot modulus $M$, total density $\rho$, fluid density $\rho_f$, apparent mass density $\rho_a$, permeability coefficient $\kappa$, porosity $\phi$, and the Biot effective stress coefficient $\alpha$. With reference to Fig.~\ref{problem_statement1}~(a), it is assumed that the plane-strain approximation holds and that the system's response to a fluid body force at $\bx_\circ = (x_\circ, y_\circ)$ is reconstructed from far-field data (or directly measured using for example convection MRI~\cite{sala2022,walk2018}) in the neighborhood $\mathscr{B}_\circ$ of the source location in terms of the solid displacement $\boldsymbol{u} = \boldsymbol{u}(\bx,\omega)$ and pore pressure $p = p(\bx,\omega)$. On recalling that $\omega$ denotes the frequency of wave motion, Biot equations~\citep{biot1962,ding2013,pour2022} may be used to describe $[\bu, p](\bx,\omega)$ for $\bx \in \mathscr{B}_\circ$ as    
\begin{equation}
\begin{aligned}
\label{eq:Biot}
& {\Uppi}_{1}[\bu, p; \delta,\omega] ~:=~ \nabla \cdot(\boldsymbol{C}\nxs:\!\nabla \boldsymbol{u}) \,-\, a \nabla p \,+\, \omega^{2} b \exs \boldsymbol{u} \,-\, \boldsymbol{f}^{u}(\delta) ~=~ \bzero, \\*[1 mm]
& {\Uppi}_{2}[\bu, p; \delta,\omega] ~:=~ \frac{c}{\omega^{2}} \nabla^{2} p \,+\, M^{-1} p \,+\, a \nabla \cdot \boldsymbol{u} \,+\, \frac{c}{\omega^{2}} f^{p}(\delta)  ~=~ 0,
\end{aligned}
\end{equation}
where $\boldsymbol{C}=\lambda \boldsymbol{I}_2 \otimes \boldsymbol{I}_2+2 \mu \boldsymbol{I}_4$ is the fourth-order drained elasticity tensor with $\boldsymbol{I}_m (m=2,4)$ representing $m$th-order symmetric identity tensor, and
\begin{equation}
\begin{aligned}
\label{eq:a}
& \gamma~=~\frac{\rho_a}{\phi^2}\,+\,\frac{\rho_f}{\phi}\,+\,\frac{\mathrm{i}}{\omega \kappa}, \quad a ~=~ \alpha \,-\, \frac{\rho_f}{\gamma}, \quad b ~=~ \rho \,-\, \frac{\rho^2_f}{\gamma}, \quad c ~=~ \frac{1}{\gamma},  \\*[1 mm]
& \boldsymbol{f}^{u} = \Big{[}- \frac{\rho_f}{\gamma} \delta(\bx - \bx_\circ), \,\, 0 \Big{]}, \quad f^p = \nabla \cdot \delta(\bx - \bx_\circ), \quad \delta(\bx - \bx_\circ) ~=~ \text{D} e^{- \varsigma \norms{ \exs \bx - \bx_\circ }^2},
\end{aligned}
\end{equation}
wherein $\delta(\bx - \bx_\circ)$ is a time-harmonic fluid body source applied at $\bx_\circ$ characterized by its magnitude $\text{D}$ and spatial decay rate $\varsigma$. \emph{Given the response $[\bu, p](\bx,\omega)$ at a fixed frequency in the neighborhood $\mathcal{B}_\circ$, the objective is to reconstruct the domain's hydromechanical properties i.e.,~$\{\mu_\circ,~\lambda_\circ,~M_\circ,~\alpha_\circ,~\phi_\circ,~\kappa_\circ \}$ in the vicinity of source point $\bx_\circ$.} It should be mentioned that the source terms $\boldsymbol{f}^{u}$ and $f^p$ in~\eqref{eq:Biot} play a key role in breaking the intrinsic symmetry in Biot equations and thus enable concurrent identification of all poroelastic parameters by eliminating the null space in the governing PDE system for every $\bx \in \mathcal{B}_\circ$. Moreover, these terms locally enhance the observability of hydraulic transport in $\mathcal{B}_\circ$ and thus improve the robustness when retrieving the permeability coefficient. The main challenge in practice pertains to characterization of tight or low-permeability formations. As such, here we assume that $\mathcal{B}$ is comprised of Pecos sandstone according to~\cite{ding2013,yew1976}. It should be mentioned that in our numerical experiments we assumed that the poroelastic properties are homogenized or constant in the vicinity of each source i.e., focal point. In what follows, the reconstructions are performed in two focal areas whose affiliated properties are listed in~Table~\ref{prop}.

\vspace*{-1mm}
\subsection{Dimensional platform}
To facilitate data processing, we normalize the system by choosing $\rho_r = 10^3$ kg/m$^3$, $\ell_r = 0.14$ m, and $\mu_r = 5.85$ GPA as the reference scales respectively for mass density, length, and stress. Note that here $\rho_r$ is identified by the fluid density, $\ell_r$ is the drained shear wavelength, and $\mu_r$ is the drained shear modulus. In this setting, all quantities are non-dimensionalized as reported in~Table~\ref{prop}. Given the multiphasic nature of the domain, the normal system retains its multiscale character and further processing is required to unify the scale of all variables. This will be accomplished by the proposed network scaling in this study.

 \begin{table}[!h]
\vspace*{-1mm}   
 \begin{center}
 \caption{\small Poroelastic properties of the domain $\mathcal{B}$ in the vicinity of two focal points.} \vspace*{-2.5mm}
\label{prop}
 \begin{tabular}{|l|l|l|}
\hline
\begin{minipage}[c][5.5mm][t]{0.1mm}%
\end{minipage}
\!\!{\small{\bf property}}  & {\small{\bf value in focal area $i = 1,2$}} & {\small{\bf dimensionless values}} 
\\  \hline \hline 
\begin{minipage}[c][5.5mm][t]{0.1mm}%
\end{minipage} 
\!\!{\small{first Lam\'{e} parameter (drained)}}  & $\lambda_1'= \lambda_2'=$ 2.74 GPA & $\lambda_1= \lambda_2=$ 0.47 
\\ \hline
\begin{minipage}[c][5.5mm][t]{0.1mm}%
\end{minipage}
\!\!{\small{drained shear modulus}}  & $\mu_1'=\mu_2'=$ 5.85  GPA & $\mu_1=\mu_2=$ 1 
\\ \hline
\begin{minipage}[c][5.5mm][t]{0.1mm}%
\end{minipage}
\!\!{\small{Biot modulus}}  & $M_1'=M_2'=$ 9.71 GPA & $M_1=M_2=$ 1.66 
\\ \hline
\begin{minipage}[c][5.5mm][t]{0.1mm}%
\end{minipage}
\!\!{\small{total density}}  & $\rho_1'=\rho_2'=$  2270 kg/m$^3$ & $\rho_1=\rho_2=$ 2.27
\\ \hline
\begin{minipage}[c][5.5mm][t]{0.1mm}%
\end{minipage}
\!\!{\small{fluid density}}  & $\rho_{f_{\exs 1}}'=\rho_{f_{\exs 2}}'=$  1000 kg/m$^3$ & $\rho_{f_{\exs 1}}=\rho_{f_{\exs 2}}=$ 1
\\ \hline
\begin{minipage}[c][5.5mm][t]{0.1mm}%
\end{minipage}
\!\!{\small{apparent mass density}}  & $\rho_{a_{1}}'=\rho_{a_{2}}'=$ 117 \, kg/m$^3$ & $\rho_{a_{1}}=\rho_{a_{2}}=$ 0.117
\\ \hline
\begin{minipage}[c][5.5mm][t]{0.1mm}%
\end{minipage}
\!\!{\small{permeability coefficient}}  & $\kappa_1'=$  503, $\kappa_2'=$  0.8  mm$^\text{4}$/N & $\kappa_1=$ 1.5407 $\!\times\!$ 10$^{-\text{5}}$, $\kappa_2=$ 2.45 $\!\times\!$ 10$^{-\text{8}}$
\\ \hline
\begin{minipage}[c][5.5mm][t]{0.1mm}%
\end{minipage}
\!\!{\small{porosity}}  & $\phi_1=\phi_2=$  0.195 & $\phi_1=\phi_2=$ 0.195
\\ \hline
\begin{minipage}[c][5.5mm][t]{0.1mm}%
\end{minipage}
\!\!{\small{Biot effective stress coefficient}}  & $\alpha_1=\alpha_2=$  0.83 &  $\alpha_1=\alpha_2=$ 0.83
\\ \hline
\begin{minipage}[c][5.5mm][t]{0.1mm}%
\end{minipage}
\!\!{\small{excitation frequency}}  & $\omega'=$  1.2 MHz& $\omega =$ 391
\\ \hline
\begin{minipage}[c][5.5mm][t]{0.1mm}%
\end{minipage}
\!\!{\small{source amplitude}}  & $\text{D}'=$  8.36 $\!\times\!$ 10$^{\text{4}}$ m& $\text{D} =$ 5.97 $\!\times\!$ 10$^{\text{5}}$
\\ \hline
\begin{minipage}[c][5.5mm][t]{0.1mm}%
\end{minipage}
\!\!{\small{source decay rate}}  & $\varsigma'=$  9.57 $\!\times\!$ 10$^{\text{4}}$ m$^{-2}$& $\varsigma =$ 187.52 
\\ \hline
\end{tabular}
\end{center}
\vspace*{-5.0mm}
\end{table}

\vspace*{-1mm}
\subsection{Data inversion}

\begin{figure}[!tp]
\begin{centering}
\includegraphics[width=0.65\columnwidth]{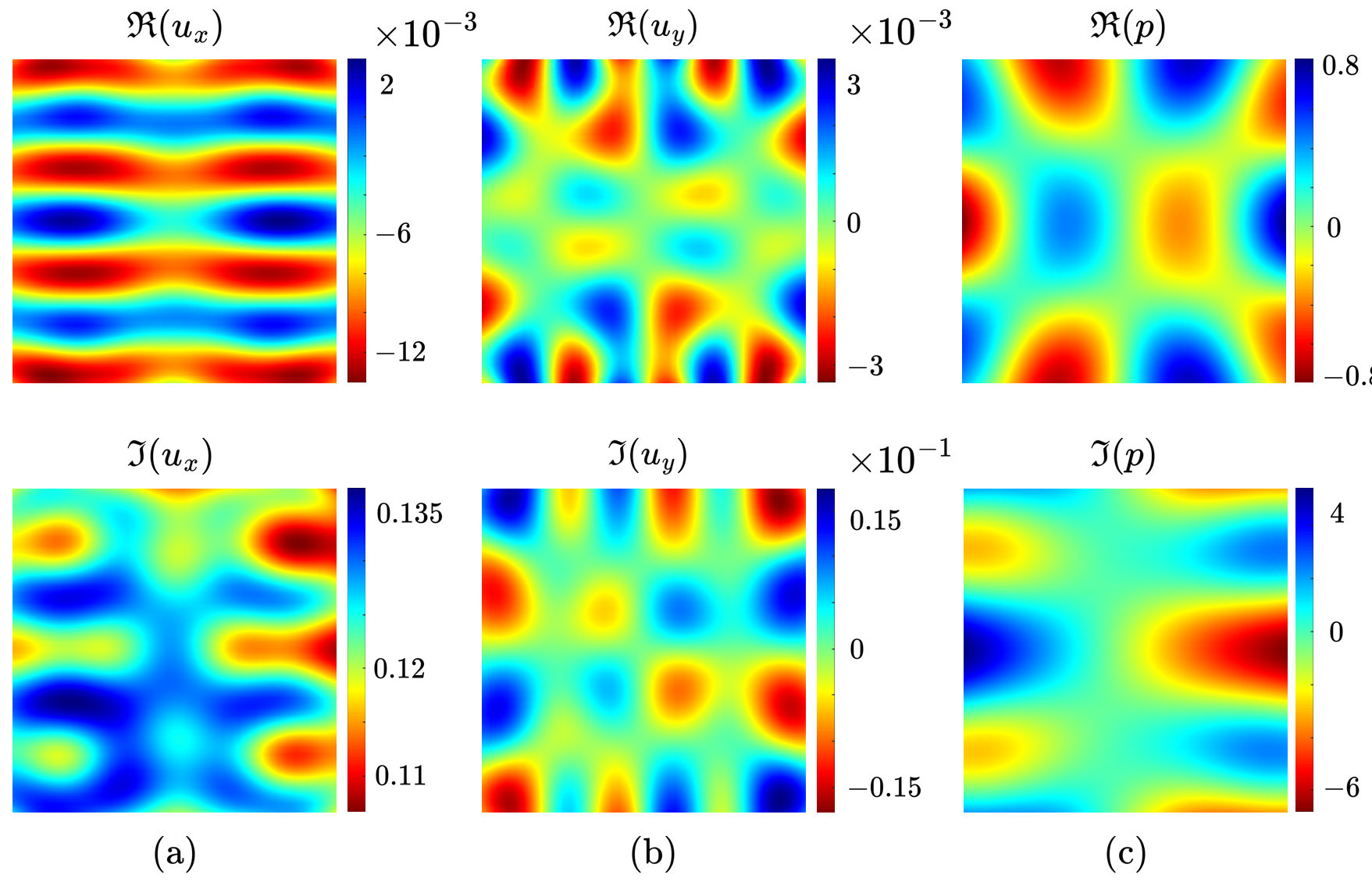}\vspace*{-2.5mm}
\par\end{centering}
\caption{Simulated poroelastic response to the excitation shown in Fig.~\ref{problem_statement1} in focal area $\bxi_1$:~real (top raw) and imaginary (bottom row) of~(a) solid displacement $u_x$, (b) solid displacement $u_y$, and~(c) interstitial pore pressure $p$.}
\label{problem_statement2}\vspace*{-4mm}
\end{figure}

In every focal area, given the frequency $\omega$ and the fluid body source $\delta(\bx-\bx_\circ)$ -- and thus, $\boldsymbol{f}^{u}(\delta)$ and $f^{p}(\delta)$ as  in Fig.~\ref{problem_statement1} -- the hydromechanical response is obtained in terms of the two-dimensional solid displacement $\bu = (u_x, u_y)$ and pore pressure $p$ as illustrated in Fig.~\ref{problem_statement2}. In this section, given the dataset $[\bu_i^j, p_i^j; \delta_i,\omega]$ in two focal regions $i = 1,2$ with $j = 1,\ldots, N_{p_i}\!\!$ data points in each neighborhood, the objective is to recover the affiliated hydromechanical properties $\{\mu_i,~\lambda_i,~M_i,~\alpha_i,~\phi_i,~\kappa_i \}$ in each region. For this purpose, the real and imaginary parts of the Biot equations as well as every complex-valued quantity are separated such that~\eqref{eq:Biot} is transformed into a real-valued system of six PDEs. The real and imaginary parts of each quantity are denoted by $\mathfrak{R}(\cdot)$ and $\mathfrak{I}(\cdot)$ respectively. The six PDE parameters in each neighborhood are mapped by a scaled MLP that is identified by minimizing the weighted residuals of the Biot system. In what follows, the property map is constructed and trained using three adaptive methods for balancing the six components of the loss function, namely: the proposed dynamic scaling, GradNorm and SoftAdapt. Each method is briefly introduced followed by a comparative analysis of their logic and performance.

\vspace*{-1mm}
\subsubsection*{Scaled neural networks as property maps}
\begin{figure}[!tp]
\begin{centering}
\includegraphics[width=0.885\columnwidth]{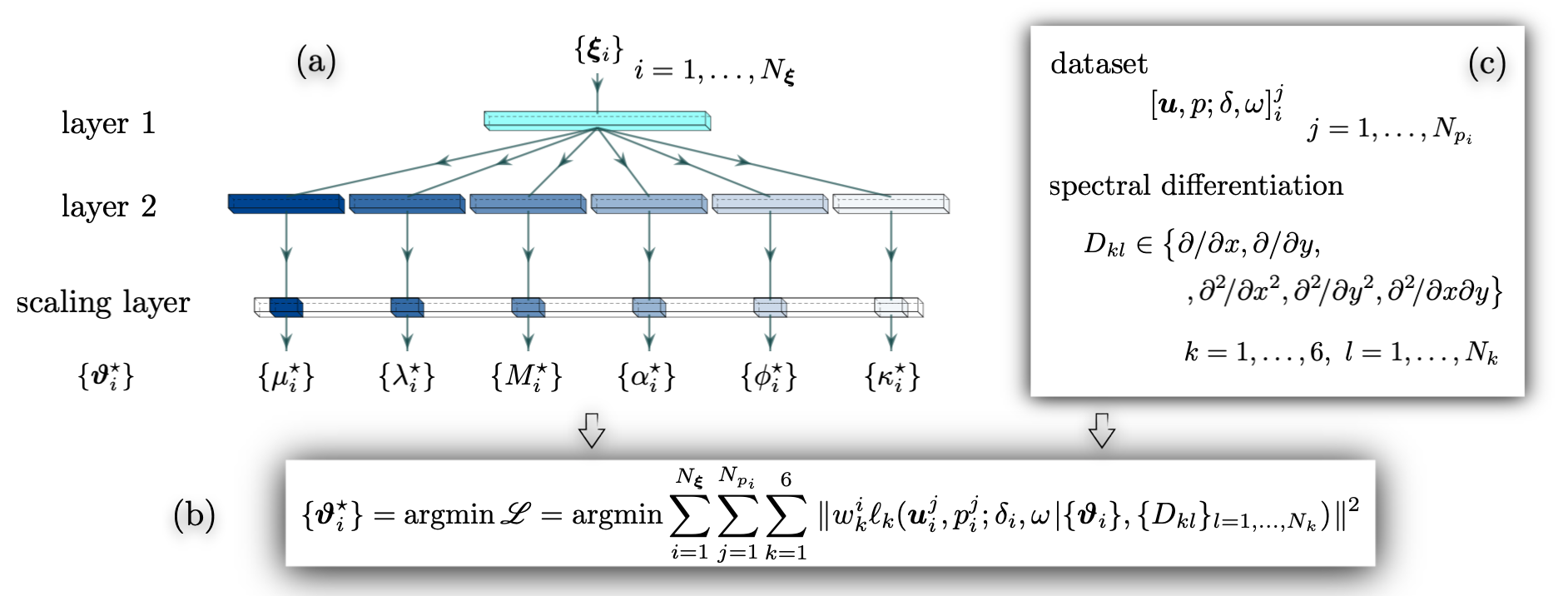}\vspace*{-2mm}
\par\end{centering}
\caption{Application of the proposed approach for intelligent poroelastography:~(a) mapping of the unknown hydromechanical properties in each focal neighborhood $\bxi_i$, $i = 1,2$, by a multiscale MLP such that the network parameters (i.e., weights and biases) remain of O(1) while the outputs are properly scaled by the last layer, (b) identifying the network by minimizing the multi-objective loss function $\mathcal{L}$ comprised of weighted PDE residuals associated with the Biot equations, and (c) the training dataset entailing the harmonic source function $\delta_i$ of frequency $\omega$ in each focal region and the associated displacement and pressure fields $[\bu_i^j, p_i^j]$ at $j = 1,\ldots, N_{p_i}$ points in the vicinity of each source along with their spatial derivatives $D_{kl}$.}\vspace*{-4mm}
\label{direct_inversion1}
\end{figure}

In this work, the unknown hydromechanical properties in the designated focal areas are mapped by a multilayer perceptron whose dense range allows for capturing arbitrary complex functions~\citep{horn1991}. As depicted in Fig.~\ref{direct_inversion1}, the input consists of a set of neighborhoods $\{\boldsymbol{\xi}_i\}$, $i = 1,\ldots, N_\xi$, here $N_\xi = 2$, while the output denotes the associated poroelastic properties. Every MLP layer (before the last one) is defined by~\eqref{eq:layer}, while the last scaling layer is defined by 
\begin{equation}
\label{eq:last_layer2}
\begin{aligned}
\big{(} \boldsymbol{\vartheta}^\star_1, \boldsymbol{\vartheta}^\star_2 \big{)} ~=~ \big{(}\{\mu^\star_1,&~\lambda^\star_1,~M^\star_1,~\alpha^\star_1,~\phi^\star_1,~\kappa^\star_1 \},  \{\mu^\star_2,~\lambda^\star_2,~M^\star_2,~\alpha^\star_2,~\phi^\star_2,~\kappa^\star_2 \} \big{)}~=~ \\*[1 mm] 
&\text{diag}\big{[} 
1,~1,~1,~1,~0.1,~s^1_\kappa, 1,~1,~1,~1,~0.1,~s^2_\kappa \big{]} \boldsymbol{x}^{N_\ell - 1}, \quad s^i_\kappa \in \{ 10^{-5}, 10^{-6}, 10^{-7}, 10^{-8} \}, \quad i = 1,2, 
\vspace*{-2mm}
\end{aligned}
\end{equation} 
in this application based on the nominal (normalized) values for sandstone. The network is trained by minimizing the loss function $\mathcal{L}$ as follows  

\begin{equation}
\label{eq:lossfunction}
 \mathcal{L} ~=~ \sum_{i= 1}^{2} \sum_{j = 1}^{N_{p_i}} \sum_{k = 1}^{6} \left\| w^i_k \ell_k\!\left(\bu_i^j, p_i^j; \delta_i, \omega \exs \big{|}  \exs\boldsymbol{\vartheta}^\star_i, \{D_{kl} \}_{l = 1,\ldots, N_k}\right) \right\|^2 ,
\end{equation}
where $\boldsymbol{\vartheta}^\star_i = (\mu^\star_i,~\lambda^\star_i,~M^\star_i,~\alpha^\star_i,~\phi^\star_i,~\kappa^\star_i )$ is the set of network-predicted properties in each focal region $\bxi_i$; $D_{kl} \in \{ \partial/\partial x, \partial/\partial y, \partial^2\!/\partial x^2, \partial^2\!/\partial y^2, \partial^2\!/\partial x \partial y \}$ is the set of differential operators according to~\eqref{Fact}, and each loss component is  
\begin{equation}
\label{eq:l1}
\{\ell_1,\exs \ell_2,\exs \ell_3,\exs \ell_4,\exs \ell_5,\exs \ell_6 \} ~=~ \big{\{}\mathfrak{R}(\Uppi_1)_x, \exs  \mathfrak{I}(\Uppi_1)_x, \exs \mathfrak{R}(\Uppi_1)_y, \exs \mathfrak{I}(\Uppi_1)_y, \exs \mathfrak{R}(\Uppi_2), \exs \mathfrak{I}(\Uppi_2) \big{\}},
\end{equation}
wherein $\Uppi_1$ and $\Uppi_2$ are defined in~\eqref{eq:Biot}. Thereby, $\ell_1$ can be written as
\begin{equation}
\label{eq:l1_expanded}
\begin{split}
 \ell_1 ~=~  &\mu\left(\frac{\partial^2 \mathfrak{R}(u_x)}{\partial x^2}\,+\,\frac{\partial^2 \mathfrak{R}(u_x)}{\partial y^2} \,+\, \frac{\partial^2 \mathfrak{R}(u_x)}{\partial x^2}\,+\,\frac{\partial^2 \mathfrak{R}(u_y)}{\partial x \partial y}\right)\,+\,\lambda\left(\frac{\partial^2 \mathfrak{R}(u_x)}{\partial x^2}\,+\,\frac{\partial^2 \mathfrak{R}(u_y)}{\partial x \partial y}\right) \,- \\
& \quad -\mathfrak{R}(a) \frac{\partial \mathfrak{R}(p)}{\partial x}\,+\,\mathfrak{I}(a) \frac{\partial \mathfrak{I}(p)}{\partial x}\,+\,\omega^2 \mathfrak{R}(b) \mathfrak{R}(u_x)\,-\,\omega^2 \mathfrak{I}(b) \mathfrak{I}(u_x) \,-\, \mathfrak{R}(f^{u}_x),
\end{split}
\end{equation}
and $\ell_2-\ell_6$ are provided in~\ref{apx} for completeness. 

\vspace*{-1mm}
\subsubsection*{Adaptive weights for loss balancing}
In this section, we briefly describe implementation of the proposed Dynamic Scaling for adaptive loss balancing, and provide an overview of the SoftAdapt and GradNorm methods. This is followed by a comparative analysis of these approaches and reporting of the obtained results in Section~\ref{ER}.  

\vspace*{-1mm}
\subsubsection*{Dynamic Scaling (DynScl)}

According to~\eqref{Fact}-\eqref{eq:dynamicweights}, this approach weights each task $\ell_k$, $k = 1,\ldots,6$, such that the affiliated residue is normalized by the average scale in $\ell_k$. For example, the weight $w_1^i$ associated with $\ell_1$ in~\eqref{eq:l1_expanded} in the $i^{\text{th}}$ neighborhood is specified as follows,  
\begin{equation}
\label{eq:dynamicweights2}
w_1^i ~=~ 10^{\exs - \beta_{1}^i}, \quad \beta_{1}^i ~=\,  \displaystyle \frac{1}{7} \sum_{l=1}^{7}\text{round}\left(\text{log}_{10}\big{[}f_{1l}\left(\boldsymbol{\vartheta}^\star_i,\omega\right)\!\big{]}\right) \,+\, \text{round}\left(\text{log}_{10}\big{[}\big{\langle} | D_{1l}[u_x,u_y,p;\boldsymbol{f}^u\nxs,f^p](\bx_i^j) | \big{\rangle}_{\nxs j}\big{]}\right), \quad i =1,2.
\end{equation}
Here, $\langle | \cdot | \rangle_{\nxs j}$ indicates the mean of absolute value of its argument over $j = 1,\ldots,N_{p_i}$ points in every focal region $\bxi_i$, and $\bx_i^j$ denotes the position vector affiliated with data point $(\bu_i^j, p_i^j)$ where $i = 1,2$. Note that all derivatives $D_{1l}[u_x,u_y,p;\boldsymbol{f}^u\nxs,f^p]$, $l = 1,\ldots,7$, are computed by way of spectral differentiation in each neighborhood from the set $\{(\bu_i^j, p_i^j,\delta_i,\omega)\}_{j = 1,\ldots,N_{p_i}}\!$. For completeness, the explicit form of operators $f_{1l}$ and $D_{1l}$, $l = 1,\ldots,7$, according to~\eqref{Fact} is provided in the following.    
\begin{equation}
\label{eq:g_11}
\begin{aligned}
&\big{\{} f_{11},~f_{12},~f_{13},~f_{14},~f_{15},~f_{16},~f_{17} \big{\}}\!\nxs \left(\boldsymbol{\vartheta}^\star_i,\omega\right)  \,=\, \big{\{} \mu^\star_i,\,\lambda^\star_i,\,-\mathfrak{R}(a^\star_i),\,\mathfrak{I}(a^\star_i),\,\omega^2  \mathfrak{R}(b^\star_i),\,-\omega^2 \mathfrak{I}(b^\star_i),\, 1\big{\}}, \\*[1 mm]
&\hspace{3cm} a^\star_i ~=~ a(\alpha^\star_i,\phi^\star_i,\kappa^\star_i), \quad b^\star_i ~=~ b(\phi^\star_i,\kappa^\star_i), \quad i =1,2, \\*[-5 mm]
\end{aligned}
\end{equation}

\begin{equation}
\label{eq:g_12}
\begin{aligned}
&\hspace{1.6cm}{\{} D_{11},~D_{12},~D_{13},~D_{14},~D_{15},~D_{16},~D_{17} {\}}\!\nxs [u_x,u_y,p;\boldsymbol{f}^u\nxs,f^p]  \,=\, \\*[1 mm]
& \left\{ \left(\frac{\partial^2 \mathfrak{R}(u_x)}{\partial x^2}+\frac{\partial^2 \mathfrak{R}(u_x)}{\partial y^2} + \frac{\partial^2 \mathfrak{R}(u_x)}{\partial x^2}+\frac{\partial^2 \mathfrak{R}(u_y)}{\partial x \partial y}\right),\left(\frac{\partial^2 \mathfrak{R}(u_x)}{\partial x^2}+\frac{\partial^2 \mathfrak{R}(u_y)}{\partial x \partial y}\right),\frac{\partial \mathfrak{R}(p)}{\partial x},\frac{\partial \mathfrak{I}(p)}{\partial x},\mathfrak{R}(u_x),\mathfrak{I}(u_x),- \mathfrak{R}(f^{u}_x) \right\}.
\end{aligned}
\end{equation}

Similar factorizations can be formulated for $\ell_2-\ell_6$ in~\eqref{eq:l3} to find $w^i_2-w^i_6$ in~\eqref{eq:lossfunction} for every focal region $\bxi_i$, $i = 1,2$. 

\vspace*{-1mm}
\subsubsection*{Softmax adaptive weights (SoftAdapt)}
In this approach, the weight of each loss component is calculated based on its live performance statistics. More specifically, the rate of change of each objective function relative to others is used to gauge its visibility to the minimizer such that the convergence of all loss components in the parameter space is approximately isotropic~\cite{heyd2019}. To this end, recall that the sought-for poroelastic properties $ \boldsymbol{\vartheta}^\star_i$ in each focal area $\bxi_i$, $i = 1,2$, is a function of network parameters i.e., weights and biases $\lbrace \boldsymbol{W}^m, \boldsymbol{b}^m \rbrace_{m = 1,2, \ldots, N_m}$ in $N_m$ layers, which in turn are a function of iteration steps $t_{\text{{n}}}$. Then, define the rate of change of the loss component $\ell_k = \ell_k\!\left(\boldsymbol{\vartheta}^\star_i(t_{\text{{n}}})\right)$ by
\begin{equation}
\label{eq:softadapt_ev}
s^i_k~=~\ell_k\!\left(\boldsymbol{\vartheta}^\star_i(t_{\text{{n}}})\right) \,-\,\, \ell_k\!\left(\boldsymbol{\vartheta}^\star_i(t_{\text{{n-1}}})\right), \quad k ~=~ 1,\ldots, 6, \quad i ~=~ 1,2, \quad {\text{{n}}} ~=~ 1,\ldots, N_{\text{{epoch}}},
\end{equation}
wherein $N_{\text{{epoch}}}$ is the number of epochs. In this setting, the loss weights are defined by
\begin{equation}
\label{eq:softadapt}
w_k^i~=~\displaystyle \frac{\text{e}^{\upeta \exs \left(s^i_k-\max\left(\{s^i_k\}_{k = 1,\ldots,6}\right)\right)}}{\sum_{k=1}^6 \text{e}^{\upeta \exs \left(s^i_k-\max\left(\{s^i_k\}_{k = 1,\ldots,6}\right)\right)}}, \quad i = 1,2,
\end{equation}
where $\upeta$ is a tunable hyper-parameter with default value of $\upeta = 0.1$. SoftAdapt assigns larger weights to loss components with slower convergence rate.

\vspace*{-1mm}
\subsubsection*{Gradient normalization (GradNorm)}
GradNorm balances the \emph{gradient of weighted loss components} with respect to network parameters to ensure that all objectives train at similar rates~\cite{chen2018}. This method addresses the common issue of gradient imbalances by penalizing tasks with excessively large or small gradients. Let $\text{\bf w} \subset \lbrace \boldsymbol{W}^m, \boldsymbol{b}^m \rbrace_{m = 1,2, \ldots, N_m}$ denote the parameters of shared MLP layer(s) in the property map, see Fig.~\ref{direct_inversion1}. GradNorm finds the optimal weights $w^i_k(t_{\text{{n}}})$, at every epoch $t_{\text{{n}}}$, by minimizing the $L_1$ difference between the actual and average norms of loss gradients with respect to $\text{\bf w}$ as the following
\begin{equation}
\label{eq:gradnorm}
\displaystyle \mathcal{L}^i_{\nabla}~=\,\,\displaystyle \sum_{k=1}^{6}\left|\left\|\nabla_{\text{\bf w}} w^i_k(t_{\text{{n}}}) \ell_k\!\left(\boldsymbol{\vartheta}^\star_i(t_{\text{{n}}})\right)\right\Vert_2\,-\,\frac{1}{6}\sum_{k=1}^{6} \left(r_k(t_{\text{{n}}})\right)^{\tilde{\upeta}} \left\|\nabla_{\text{\bf w}} w^i_k(t_{\text{{n}}}) \ell_k\!\left(\boldsymbol{\vartheta}^\star_i(t_{\text{{n}}})\right)\right\|_2 \right|_1, \quad  r_k(t_{\text{{n}}})~=~\frac{{\ell_k\!\left(t_{\text{{n}}}\right)}/{\ell_k\!\left(t_{1}\right)}}{ \sum_{k=1}^{6}{\ell_k\!\left(t_{\text{{n}}}\right)}/{\ell_k\!\left(t_{1}\right)}},
\end{equation}
where $\tilde{\upeta}$ is a tunable hyperparameter that determines the strength of the restoring force that aligns the loss components to a common convergence rate. When loss components differ greatly in scale, causing significant variations in the loss distribution during training, a higher value of $\tilde{\upeta}$ is needed to enforce a balance. In contrast, when the scale of loss component are more similar, a lower value of $\tilde{\upeta}$ is suitable. Taking $\tilde{\upeta}=0$ will assign equal weights to all components. Note that minimizing $\mathcal{L}^i_{\nabla}$ is repeated at every epoch while optimizing $\mathcal{L}$. Since $w^i_k$ directly influences the gradient magnitudes for each loss component, $\mathcal{L}^i_{\nabla}$ is differentiated with respect to $w^i_k$, and the resulting derivatives are used to update each weight $w^i_k$ via standard gradient decent approaches.

\vspace*{-1mm}
\subsubsection*{Comparative analysis}

The proposed approach based on dynamic scaling of the loss components is closely related to GradNorm. To observe this, recall that the weights and biases remain of O(1) at every epoch in the property map and thus
\begin{equation}
\label{Nds}
\frac{\partial{\left({\vartheta}^\star_i\right)}_{\nxs m}}{\partial \text{w}_{n}} ~=~ O\left({\left(s_m\right)}_{i}\right), \quad m = 1,\ldots, 6, \,\,\, n = 1,\ldots,N_{\text{\bf w}}, \,\,\, i = 1,2, \quad {\left({\vartheta}^\star_i\right)}_{\nxs m} \in \boldsymbol{\vartheta}^\star_i, \,\,\, \text{w}_{n} \in \text{\bf w},
\end{equation}
wherein ${(s_m)}_{i}$ is the relevant scale in the MLP's last layer. Whereby, it is straightforward from~\eqref{eq:dynamicweights2}-\eqref{eq:g_12} to prove that 
\begin{equation}
\label{DS-GN}
\begin{aligned}
\frac{\partial w^i_1(t_{\text{{n}}}) \ell_1\!\left(\boldsymbol{\vartheta}^\star_i(t_{\text{{n}}})\right)}{\partial \text{w}_{n}} &~=~ \sum_{l=1}^{7} w^i_1(t_{\text{{n}}}) D_{1l}[u_x,u_y,p;\boldsymbol{f}^u\nxs,f^p]  \frac{\partial f_{1l}(\boldsymbol{\vartheta}^\star_i,\omega)}{\partial \text{w}_{n}} ~=~ \\*[1mm]
&~=~ \sum_{l=1}^{7} w^i_1(t_{\text{{n}}}) D_{1l}[u_x,u_y,p;\boldsymbol{f}^u\nxs,f^p] \nabla_{\boldsymbol{\vartheta}^\star_i} f_{1l}(\boldsymbol{\vartheta}^\star_i,\omega)  \frac{\partial \boldsymbol{\vartheta}^\star_i}{\partial \text{w}_{n}} ~=~ O(1), \quad n = 1,\ldots,N_{\text{\bf w}}, \,\,\, i = 1,2.
\end{aligned}
\end{equation}

This remains the case for all other weighted components of the loss $w^i_k(t_{\text{{n}}}) \ell_k\!\left(\boldsymbol{\vartheta}^\star_i(t_{\text{{n}}})\right)$, for $k = 2,\ldots, 6$, as one may prove that $f_{kl}(\boldsymbol{\vartheta}^\star_i,\omega)$, corresponding to the factorization in~\eqref{Fact} for the Biot system, are Lipchitz continuous with respect to $\text{w}_{n} \in \text{\bf w}$ such that the Lipchitz constant is of $O\left(f_{kl}\right)$. In other words,       
\begin{equation}
\label{LPZ}
 \frac{\partial f_{kl}}{\partial \text{w}_{n}} ~=~ O\left(f_{kl}\right), \quad k = 2,\ldots, 6,  \quad n = 1,\ldots,N_{\text{\bf w}}, \quad i = 1,2.
\end{equation}

Therefore, the Dynamic Scaling approach automatically achieves the GradNorm's objective in~\eqref{eq:gradnorm} without requiring a separate minimization procedure per epoch. This is accomplished by \emph{approximating}, and thereby normalizing, the scale of relevant derivatives instead of their explicit calculation which may involve significant computational cost and complications during the optimization process. This solution is afforded by the proposed architecture of the property network and its explicit scaling via the last layer. Note that in many physical systems given the frequency of wave motion and network-predicted physical properties, the scale of spatial derivatives $D_{kl}$ in the PDE system can be estimated without explicit differentiation. This will be particularly helpful for verification and validation of the computed derivatives, especially in presence of noise in data, and in certain scenarios can help expedite the loss balancing process.     

Softmax adaptive weights carry the advantage of being computationally efficient and flexible so that they can be applied to a wide range of network architectures and multi-objective loss functions. There are two caveats, however, when it comes to using SoftAdapt for balancing multi-physics PDE systems: (i) the heuristic nature of the Softmax function that may not cater, in terms of its range, for systems that span several scales in time-space, and (ii) this approach uniformly assigns larger weights to loss components with lower rate of change. This is typically done without differentiating whether, for instance, the computed rate of change is large due to scaling or due to convergence. In physical systems, one objective may assume much larger values compared to others and shows a significant decay rate without actually converging, while other loss components could be fast converging but due to their smaller scale show much smaller decay rates. This could lead to training instability, especially in presence of noise.

\section{Implementation results}\label{ER}

\vspace*{-1.0mm}
\subsection{Reconstruction from noiseless data}\label{NLD}

In this section, we report the reconstruction results related to two focal regions introduced in Section~\ref{MP}. Here, the proposed network scaling is coupled with three different loss balancing techniques, described in Section~\ref{MP}, namely: the proposed dynamic scaling (DS), GradNorm (GN), and SoftAdapt (SA). We compare the performance of these methods based on  accuracy and robustness of the associated results. In all cases, the property maps are constructed by MLPs comprised of two parts; the first part is a single fully-connected tanh-activated layer with thirty two neurons as a common trunk, while the second part involves six individual towers,
\begin{figure}[bp!]
\vspace*{-2mm}
\center\includegraphics[width=0.75\columnwidth]{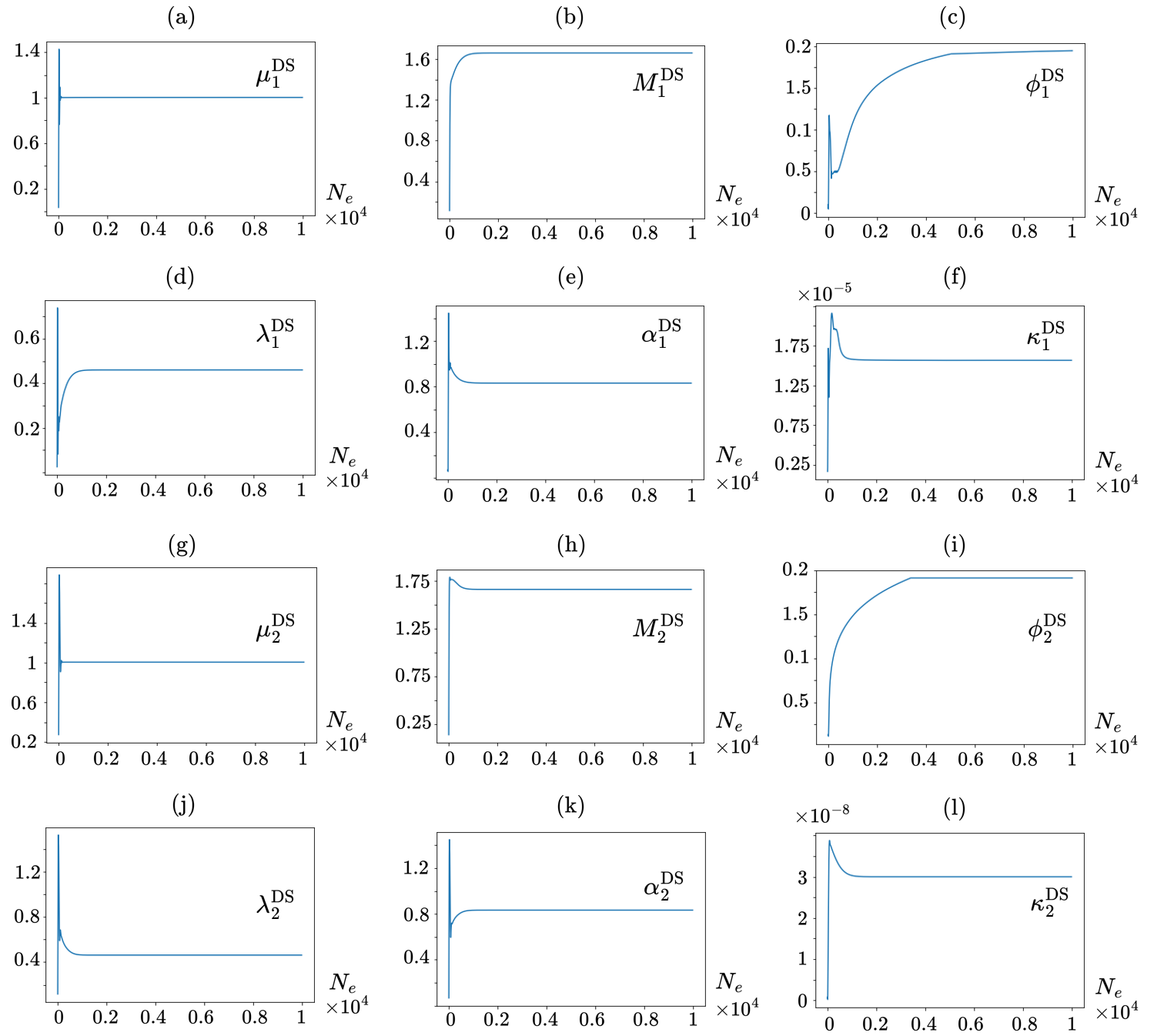}\vspace*{-2mm}
\caption{Network-predicted poroelastic properties~\emph{vs.}~number of epochs $N_{e}$ when the reconstruction is simultaneously conducted in the focal regions $\{ \bxi_i \}$, $i =1,2$. The network is a scaled MLP and the loss is balanced using the proposed Dynamic Scaling (DS) approach:~(a, g) drained shear modulus $\mu^{\text{\tiny DS}}_i$, (b, h)~Biot modulus $M^{\text{\tiny DS}}_i$, (c, i)~porosity $\phi^{\text{\tiny DS}}_i$, (d, j)~drained first Lam\'{e} parameter $\lambda^{\text{\tiny DS}}_i$, (e, k)~Biot effective stress coefficient $\alpha^{\text{\tiny DS}}_i$, (f, l)~permeability coefficient $\kappa^{\text{\tiny DS}}_i$.}\vspace*{-2mm} \label{DW_combined_k} 
\center\includegraphics[width=1\columnwidth]{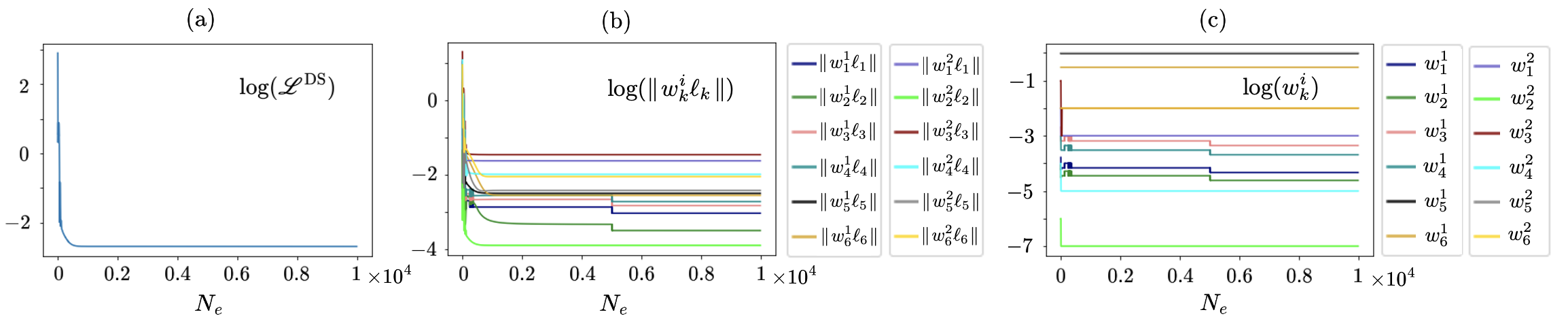}\vspace*{-4mm}
\caption{Convergence plots i.e.,~loss~\emph{vs.}~epoch corresponding to the results of Fig.~\ref{DW_combined_k} where the reconstruction is simultaneously conducted in the focal regions $\{ \bxi_i \}$, $i =1,2$. The network is a scaled MLP and the loss is balanced using the proposed Dynamic Scaling (DS) approach: (a) total loss $\text{log}(\mathscr{L}^{\text{\tiny DS}})$ trajectory against the number of epochs, (b) weighted loss components $\text{log}(\norms{w^{i}_k \ell_k})$ with $\ell_k$, $k = 1,\ldots,6$, denoting the $k^{\text{th}}$ loss component and $w^{i}_k$ the associated weight in the $i^{\text{th}}$ focal region, (c) loss weights $\text{log}(w^{k}_{i})$.}\vspace*{-4mm}
\label{DW_combined_k2}\vspace*{-2mm} 
\end{figure}
each of which composed of sixteen neurons whose output is separately scaled by the last layer to predict a designated material parameter. This architecture is consistent with the schematic shown in Fig.~\ref{direct_inversion1}. All models are trained by the Adam optimizer, and the learning rate is tuned during the training process.  We also compare the results to those obtained without network scaling and/or loss balancing. The poroelastodynamic simulations, germane to the configuration shown in Fig.~\ref{problem_statement1}~(a), are conducted using the FreeFem++~\citep{Hech2012} platform to generate the training data. In this vein, we built on an existing code that was recently developed as part of~\citep{pour2022} to model poroelastic wave motion in the subsurface. In what follows, the reported reconstruction error is normalized according to      
\begin{equation}\label{normal_misfit_parameter}
\Xi\left(({\vartheta}^\star_{i})_{n}\right) ~=~\frac{{\left|({\vartheta}^\star_{i})_{n}\,-\,({\vartheta}_{i})_{n}\right|}}{{\left|({\vartheta}_{i})_{n}\right|}}, \quad i ~=~ 1,2, \quad n ~=~ 1,\ldots, N_{\boldsymbol{\vartheta}}, \quad \star ~=~\{{\text{EW}},~{\text{DS}},~{\text{GN}},~{\text{SA}}\},
\end{equation}
where $({\vartheta}^\star_{i})_{n}$ is the neural network prediction for a quantity whose true value is $({\vartheta}_{i})_{n}$. Note that $({\vartheta}_{i})_{n}$ indicates the $n^{\text{th}}$ component of the property vector $\boldsymbol{\vartheta}_i$ in the $i^{\text{th}}$ focal region when $i = 1,2$ in this study. It should be mentioned that in this section for clarity the superscript $\star$ which indicates network prediction in Section~\ref{MP} is replaced by an abbreviation of the method used for training.

The reconstruction results using the proposed dynamic scaling approach is shown in Figs.~\ref{DW_combined_k} and~\ref{DW_combined_k2}. Note that here all data from both low-permeability and high-permeability neighborhoods $\{\bxi_i\}$, $i = 1,2$, is used for training to simultaneously reconstruct the six unknown poroelastic parameters in each neighborhood leading to twelve network outputs $( \boldsymbol{\vartheta}^{\text{\tiny DS}}_1, \boldsymbol{\vartheta}^{\text{\tiny DS}}_2 ) = (\{\mu^{\text{\tiny DS}}_1,~\lambda^{\text{\tiny DS}}_1,~M^{\text{\tiny DS}}_1,~\alpha^{\text{\tiny DS}}_1,~\phi^{\text{\tiny DS}}_1,~\kappa^{\text{\tiny DS}}_1 \},  \{\mu^{\text{\tiny DS}}_2,~\lambda^{\text{\tiny DS}}_2,~M^{\text{\tiny DS}}_2,~\alpha^{\text{\tiny DS}}_2,~\phi^{\text{\tiny DS}}_2,~\kappa^{\text{\tiny DS}}_2 \})$. Our comparative analysis with GradNorm and SoftAdapt predictions is however conducted in each neighborhood separately. This is motivated by the logical comparison of these approaches, provided at the end of Section~\ref{MP}, realizing that SoftAdapt associates the variation of loss components only to their convergence behavior and may not differentiate the impact of distinct physical scales on variation magnitudes of different objectives. Moreover, for computational efficiency, the GradNorm typically normalizes the gradients of weighted loss components with respect to the parameters in a shared layer in the neural network (as opposed to all network parameters) which may create some stiffness in loss balancing depending on the network architecture and input/output. Given this, it may be more insightful to compare the reconstruction results in each neighborhood separately. This may also shine some light on the impact of permeability on the reconstruction results using different methods.  

\begin{figure}
\begin{centering}
\includegraphics[width=1\columnwidth]{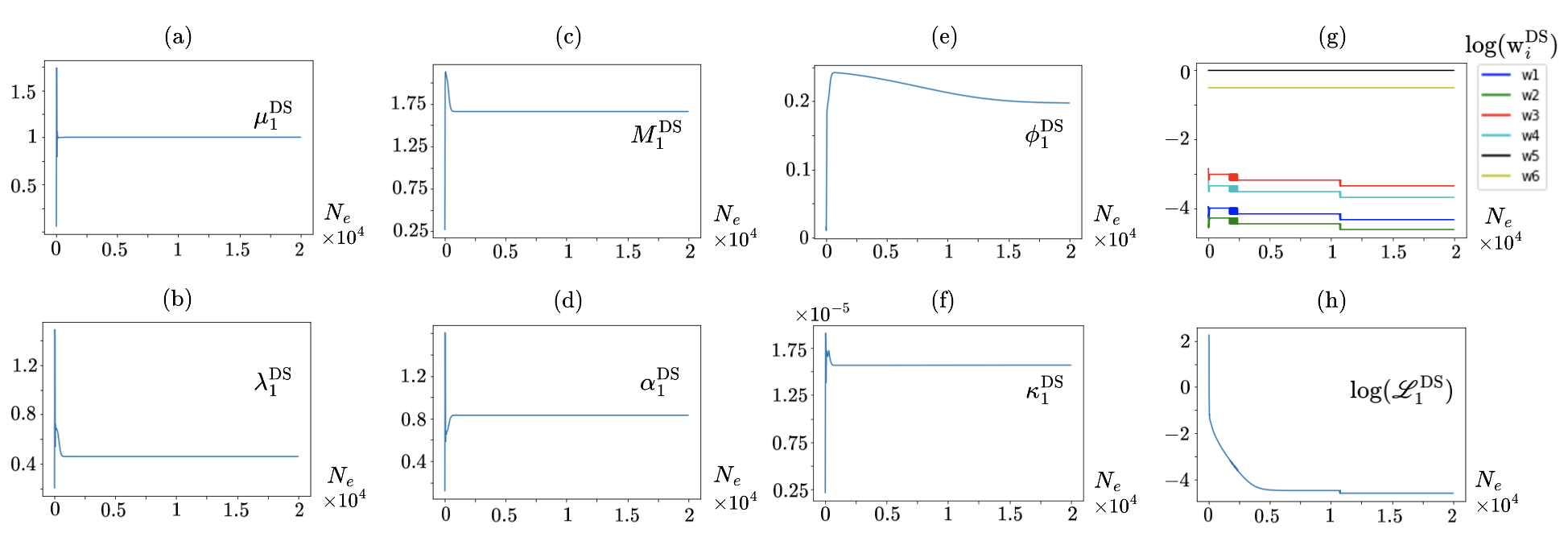}\vspace*{-2.0mm}
\par\end{centering}
\caption{Network-predicted poroelastic properties~\emph{vs.}~the number of epochs $N_{e}$ when the reconstruction is conducted in the high-permeability neighborhood $\bxi_1$. The network is a scaled MLP and the loss is balanced using the proposed Dynamic Scaling (DS) approach:~(a) drained shear modulus $\mu^{\text{\tiny DS}}_1\!$, (b)~drained first Lam\'{e} parameter $\lambda^{\text{\tiny DS}}_1\!$, (c)~Biot modulus $M^{\text{\tiny DS}}_1\!$, (d)~Biot effective stress coefficient $\alpha^{\text{\tiny DS}}_1\!$, (e)~porosity $\phi^{\text{\tiny DS}}_1\!$, (f)~permeability coefficient $\kappa^{\text{\tiny DS}}_1\!$, (g) DS weights $\text{w}^{\text{\tiny DS}}_i$, $i = 1, \ldots, 6$, versus the minimizer step $N_{e}$, (h) weighted total loss $\text{log}(\mathscr{L}_1^{\text{\tiny DS}})$ trajectory against the number of epochs.}
\label{DW_higher_k}
\vspace*{-4.0mm}
\end{figure}

\begin{figure}
\begin{centering}
\includegraphics[width=1\columnwidth]{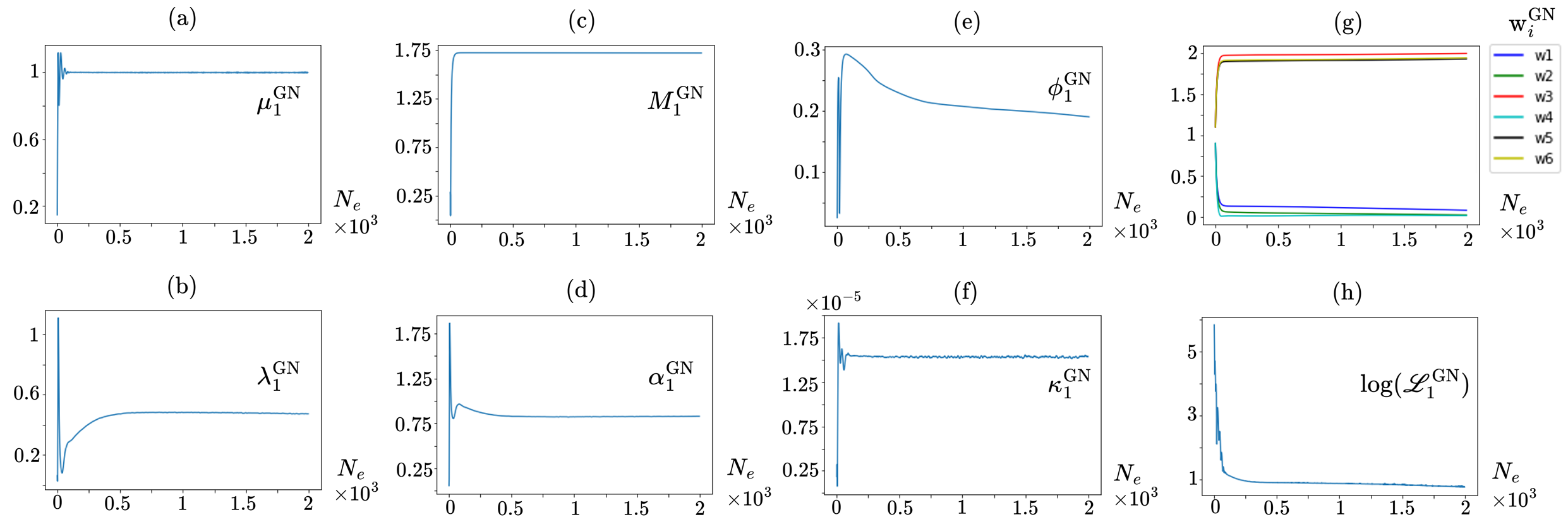}\vspace*{-2.0mm}
\par\end{centering}
\caption{Network-predicted poroelastic properties~\emph{vs.}~the number of epochs $N_{e}$ in the high-permeability neighborhood $\bxi_1$. The network is a scaled MLP and the loss is balanced using the GradNorm (GN) approach:~(a) drained shear modulus $\mu^{\text{\tiny GN}}_1\!$, (b)~drained first Lam\'{e} parameter $\lambda^{\text{\tiny GN}}_1\!$, (c)~Biot modulus $M^{\text{\tiny GN}}_1\!$, (d)~Biot effective stress coefficient $\alpha^{\text{\tiny GN}}_1\!$, (e)~porosity $\phi^{\text{\tiny GN}}_1\!$, (f)~permeability coefficient $\kappa^{\text{\tiny GN}}_1\!$, (g) GN weights $\text{w}^{\text{\tiny GN}}_i$, $i = 1, \ldots, 6$, versus the minimizer step $N_{e}$, (h) weighted total loss $\text{log}(\mathscr{L}_1^{\text{\tiny GN}})$~\emph{vs.}~epoch.}
\label{GN_higher_k}
\vspace*{-4.0mm}
\end{figure}

\begin{figure}
\begin{centering}
\includegraphics[width=1\columnwidth]{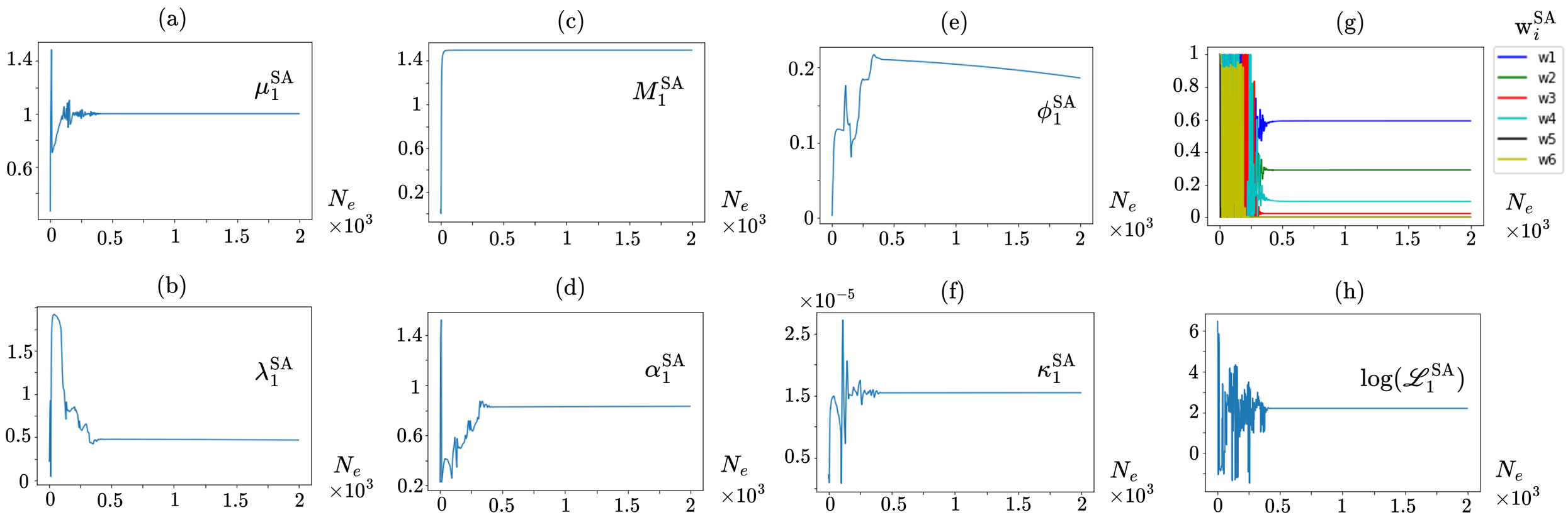}\vspace*{-2.0mm}
\par\end{centering}
\caption{Network-predicted poroelastic properties~\emph{vs.}~the number of epochs $N_{e}$ in the high-permeability neighborhood $\bxi_1$. The network is a scaled MLP and the loss is balanced using the SoftAdapt (SA) approach:~(a) drained shear modulus $\mu^{\text{\tiny SA}}_1\!$, (b)~drained first Lam\'{e} parameter $\lambda^{\text{\tiny SA}}_1\!$, (c)~Biot modulus $M^{\text{\tiny SA}}_1\!$, (d)~Biot effective stress coefficient $\alpha^{\text{\tiny SA}}_1\!$, (e)~porosity $\phi^{\text{\tiny SA}}_1\!$, (f)~permeability coefficient $\kappa^{\text{\tiny SA}}_1\!$, (g) SA weights $\text{w}^{\text{\tiny SA}}_i$, $i = 1, \ldots, 6$, versus the minimizer step $N_{e}$, (h) weighted total loss $\text{log}(\mathscr{L}_1^{\text{\tiny SA}})$~\emph{vs.}~epoch.}
\label{SA_higher_k}
\vspace*{-4.0mm}
\end{figure}

Table~\ref{gt_vs_reconstruction_NS_HK} provides the reconstruction results and associated normal errors, with respect to the ground truth, in the high-permeability neighborhood $\bxi_1$. Here, the property map is a scaled MLP as shown in Fig.~\ref{direct_inversion1}. Four loss balancing schemes are implemented, namely: DynScl (DS), GradNorm (GN), SoftAdapt (SA), and equal weights (EW) or no balancing. The reconstruction results are then accordingly denoted by $\boldsymbol{\vartheta}^{\text{\tiny DS}}_1$, $\boldsymbol{\vartheta}^{\text{\tiny GN}}_1$, $\boldsymbol{\vartheta}^{\text{\tiny SA}}_1$, and $\boldsymbol{\vartheta}^{\text{\tiny EW}}_1$ where 
$$\boldsymbol{\vartheta}^\star_1 ~=~ \{\mu^\star_1,~\lambda^\star_1,~M^\star_1,~\alpha^\star_1,~\phi^\star_1,~\kappa^\star_1 \}, \quad  \star ~=~\{{\text{EW}},~{\text{DS}},~{\text{GN}},~{\text{SA}}\}. $$
In all cases, the training data and network architecture remains the same. DS, GN, and SA methods outperform the case of equal weights, highlighting the importance of loss balancing in identification of multiphysics and multiscale systems. Note that the maximum normal error associated with EW, DS, GN, and SA reconstructions respectively read $100\%$, $2.2\%$, $18.36\%$, and $9.8\%$. The largest error does not correspond to the same parameter in different reconstructions. More specifically, the maximum error using equal weights and dynamic scaling is related to the drained first Lam\'{e} parameter $\lambda^{\text{\tiny EW}}_1\!$ and $\lambda^{\text{\tiny DS}}_1$, while in the case of GradNorm and SoftAdapt, the largest error pertains to porosity $\phi^{\text{\tiny GN}}_1\!$ and Biot modulus $M^{\text{\tiny SA}}_1\!$ respectively. The convergence plots along with network predictions as a function of optimizer step $N_e$ are shown in Figs~\ref{DW_higher_k},~\ref{GN_higher_k},~\ref{SA_higher_k} for reconstructions by way of DynScl, GradNorm, and SoftAdapt respectively.     

Table~\ref{gt_vs_reconstruction_HK} reports the reconstruction results when the exercise in Table~\ref{gt_vs_reconstruction_NS_HK} is repeated without network scaling. Here, the property map is an MLP without the scaling layer i.e., the scaling layer is replaced by an affine map with unknown parameters that is standard in MLP architectures. In this case, the reconstruction fails regardless of the choice of loss balancing technique. This exercise highlights the importance of proper scaling of the model in tandem with loss balancing for a successful data inversion. It is note worthy that the maximum normalized error is first associated with the permeability coefficient $\kappa^\star_1$ and next porosity $\phi^\star_i$ both related to the fluid dynamics in the interstitial pore space that is distinct (in terms of scale) from the physics of wave motion in the solid rock.      

\begin{figure}
\begin{centering}
\includegraphics[width=1\columnwidth]{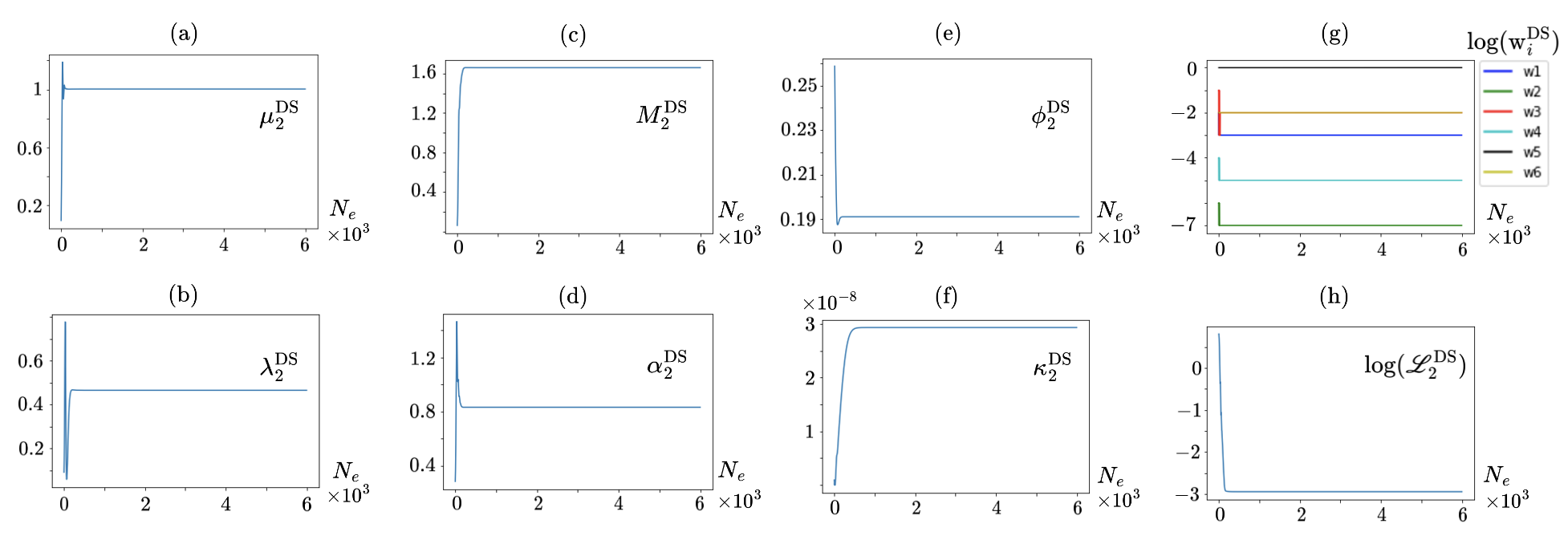}\vspace*{-2.0mm}
\par\end{centering}
\caption{Network-predicted poroelastic properties~\emph{vs.}~the number of epochs $N_{e}$ when the reconstruction is conducted in the low-permeability neighborhood $\bxi_2$. The network is a scaled MLP and the loss is balanced using the proposed Dynamic Scaling (DS) approach:~(a) drained shear modulus $\mu^{\text{\tiny DS}}_2\!$, (b)~drained first Lam\'{e} parameter $\lambda^{\text{\tiny DS}}_2\!$, (c)~Biot modulus $M^{\text{\tiny DS}}_2\!$, (d)~Biot effective stress coefficient $\alpha^{\text{\tiny DS}}_2\!$, (e)~porosity $\phi^{\text{\tiny DS}}_2\!$, (f)~permeability coefficient $\kappa^{\text{\tiny DS}}_2\!$, (g) DS weights $\text{w}^{\text{\tiny DS}}_i$, $i = 1, \ldots, 6$, versus the minimizer step $N_{e}$, (h) weighted total loss $\text{log}(\mathscr{L}_2^{\text{\tiny DS}})$ trajectory against the number of epochs.}
\label{DW_lower_k}
\vspace*{-4.0mm}
\end{figure}

\begin{figure}
\begin{centering}
\includegraphics[width=1\columnwidth]{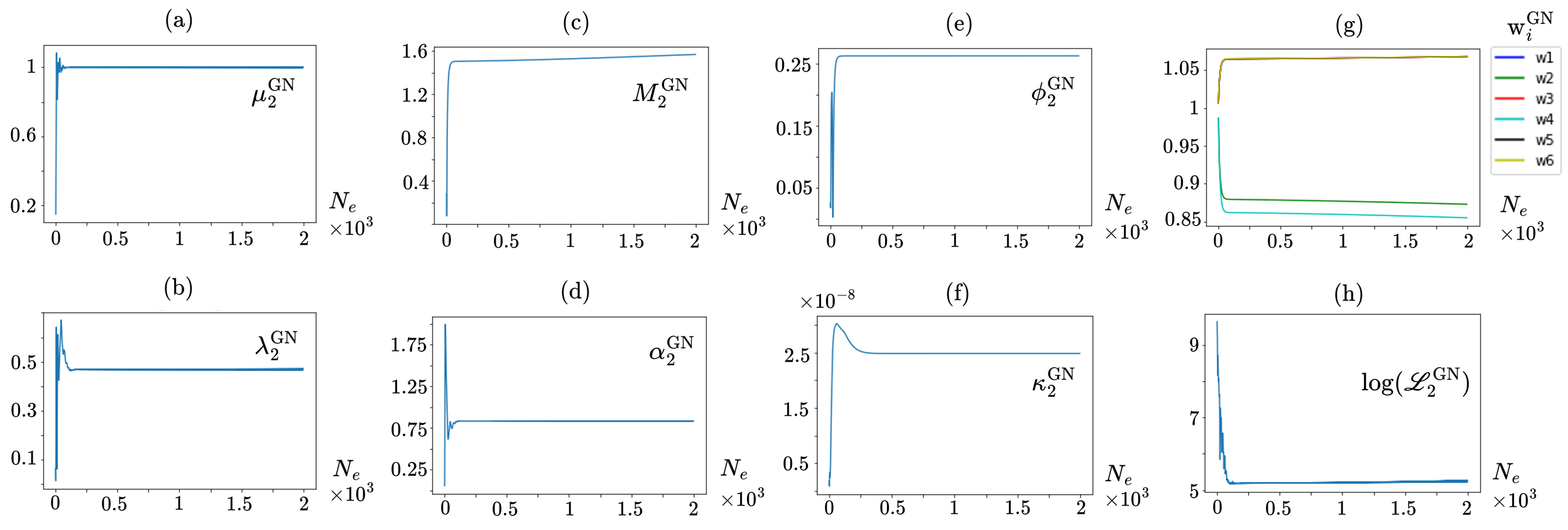}\vspace*{-2.0mm}
\par\end{centering}
\caption{Network-predicted poroelastic properties~\emph{vs.}~the number of epochs $N_{e}$ in the low-permeability neighborhood $\bxi_2$. The network is a scaled MLP and the loss is balanced using the GradNorm (GN) approach:~(a) drained shear modulus $\mu^{\text{\tiny GN}}_2\!$, (b)~drained first Lam\'{e} parameter $\lambda^{\text{\tiny GN}}_2\!$, (c)~Biot modulus $M^{\text{\tiny GN}}_2\!$, (d)~Biot effective stress coefficient $\alpha^{\text{\tiny GN}}_2\!$, (e)~porosity $\phi^{\text{\tiny GN}}_2\!$, (f)~permeability coefficient $\kappa^{\text{\tiny GN}}_2\!$, (g) GN weights $\text{w}^{\text{\tiny GN}}_i$, $i = 1, \ldots, 6$, versus the minimizer step $N_{e}$, (h) weighted total loss $\text{log}(\mathscr{L}_2^{\text{\tiny GN}})$~\emph{vs.}~epoch.}
\label{GN_lower_k}
\vspace*{-4.0mm}
\end{figure}

\begin{figure}
\begin{centering}
\includegraphics[width=1\columnwidth]{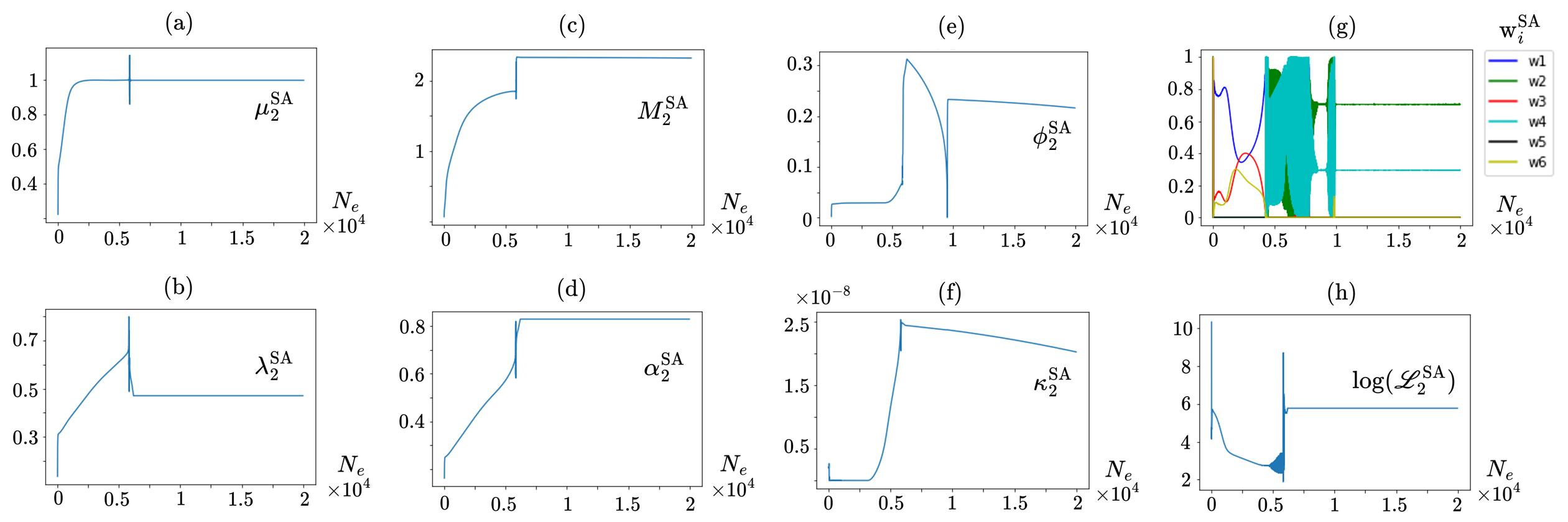}\vspace*{-2.0mm}
\par\end{centering}
\caption{Network-predicted poroelastic properties~\emph{vs.}~the number of epochs $N_{e}$ in the low-permeability neighborhood $\bxi_2$. The network is a scaled MLP and the loss is balanced using the SoftAdapt (SA) approach:~(a) drained shear modulus $\mu^{\text{\tiny SA}}_2\!$, (b)~drained first Lam\'{e} parameter $\lambda^{\text{\tiny SA}}_2\!$, (c)~Biot modulus $M^{\text{\tiny SA}}_2\!$, (d)~Biot effective stress coefficient $\alpha^{\text{\tiny SA}}_2\!$, (e)~porosity $\phi^{\text{\tiny SA}}_2\!$, (f)~permeability coefficient $\kappa^{\text{\tiny SA}}_2\!$, (g) SA weights $\text{w}^{\text{\tiny SA}}_i$, $i = 1, \ldots, 6$, versus the minimizer step $N_{e}$, (h) weighted total loss $\text{log}(\mathscr{L}_2^{\text{\tiny SA}})$~\emph{vs.}~epoch.}
\label{SA_lower_k}
\vspace*{-4.0mm}
\end{figure}

Table~\ref{gt_vs_reconstruction_NS_LK} presents the reconstruction results and affiliated normalized errors in the low-permeability neighborhood $\bxi_2$. Similar to Table~\ref{gt_vs_reconstruction_NS_HK}, the property map is a scaled MLP as in Fig.~\ref{direct_inversion1} and four loss balancing techniques are deployed for training. Again, DS, GN, and SA methods outperform the case of equal weights. The largest error using equal weights, dynamic scaling, GradNorm and SoftAdapt are respectively related to porosity $\phi^{\text{\tiny EW}}_2\!$, permeability coefficient $\kappa^{\text{\tiny DS}}_2$, porosity $\phi^{\text{\tiny GN}}_2\!$ and Biot modulus $M^{\text{\tiny SA}}_2\!$ and read $71\%$, $20\%$, $35\%$, and $40\%$. In the SoftAdapt reconstruction, there are multiple terms with large associated normal error. The convergence plots along with network predictions as a function of the number of epochs $N_e$ are shown in Figs~\ref{DW_lower_k},~\ref{GN_lower_k},~\ref{SA_lower_k} for reconstructions by way of DynScl, GradNorm, and SoftAdapt respectively.

\begin{table}[!bp]
\vspace*{-2.0mm}
\fontsize{10}{12}\selectfont \caption{Reconstructed poroelastic properties in the high-permeability neighborhood $\bxi_1$. Network scaling is applied here.}
\label{gt_vs_reconstruction_NS_HK}
\centering{}%
\vspace*{-2.0mm}
\begin{tabular}{ccccccc}
\toprule
$\boldsymbol{\vartheta}_1$ & $\mu_1$ & $\lambda_1$ & $M_1$ & $\alpha_1$  & $\phi_1$ & $\kappa_1$\\*[0.5mm]
\midrule
{\small ground truth (dimensionless value)} & 1 & 0.47 & 1.66 & 0.83 & 0.195 & $1.5407 \times 10^{-5}$\\*[0.5mm]
\midrule
{\small Equal Weights ($\boldsymbol{\vartheta}^{\text{\tiny EW}}_1$)}\!\!\!\! & 1.0003 & 0.0005 & 1.4937 & 1.1618 & 0.0492 & $1.9866 \times 10^{-5}$\\*[0.5mm]
{\small DynScl ($\boldsymbol{\vartheta}^{\text{\tiny DS}}_1$)} & 1.0015 & 0.4599 & 1.6603 & 0.8307 & 0.1971 & $1.5667 \times 10^{-5}$\\*[0.5mm]
{\small GradNorm ($\boldsymbol{\vartheta}^{\text{\tiny GN}}_1$)}  & 0.99995 & 0.4721 & 1.72156 & 0.8276 & 0.2308 & $1.5369 \times 10^{-5}$\\*[0.5mm]
{\small SoftAdapt ($\boldsymbol{\vartheta}^{\text{\tiny SA}}_1$)} & 0.9999 & 0.4665 & 1.4974 & 0.832 & 0.1861 & $1.5447 \times 10^{-5}$\\*[0.5mm]
\midrule
$\Xi$($\boldsymbol{\vartheta}^{\text{\tiny EW}}_1$) & 0.03\% & 100\% & 10\% & 40\% & 75\% & 29\%\\*[0.5mm]
$\Xi$($\boldsymbol{\vartheta}^{\text{\tiny DS}}_1$) & 0.15\% & 2.2\% & 0.018\% & 0.084\% & 1.4\% & 1.7\%\\*[0.5mm]
$\Xi$($\boldsymbol{\vartheta}^{\text{\tiny GN}}_1$) & 0.005\% & 0.45\% & 3.7\% & 0.29\% & 18.36\% & 0.25\%\\*[0.5mm]
$\Xi$($\boldsymbol{\vartheta}^{\text{\tiny SA}}_1$) & 0.01\% & 0.74\% & 9.8\% & 0.24\% & 4.6\% & 0.26\%\\*[0.5mm]
\bottomrule
\end{tabular}
\vspace*{7.0mm}
\fontsize{10}{12}\selectfont \caption{Reconstructed poroelastic properties in the high-permeability neighborhood $\bxi_1$ without network scaling.}
\label{gt_vs_reconstruction_HK}
\centering{}%
\vspace*{-2.0mm}
\begin{tabular}{ccccccc}
\toprule
$\boldsymbol{\vartheta}_1$ & $\mu_1$ & $\lambda_1$ & $M_1$ & $\alpha_1$  & $\phi_1$ & $\kappa_1$\\*[0.5mm]
\midrule
{\small ground truth} & 1 & 0.47 & 1.66 & 0.83 & 0.195 & $1.5407 \times 10^{-5}$\\*[0.5mm]
\midrule
{\small Equal Weights ($\boldsymbol{\vartheta}^{\text{\tiny EW}}_1$)}\!\!\!\! & 0.991 & 0.0006123 & 1.5014 & 1.1456 & 0.0214 & 2.0242\\*[0.5mm]
{\small DynScl ($\boldsymbol{\vartheta}^{\text{\tiny DS}}_1$)} & 0.662 & 1.813 & 1.491 & 1.617 & 0.121 & 0.987\\*[0.5mm]
{\small GradNorm ($\boldsymbol{\vartheta}^{\text{\tiny GN}}_1$)} & 0.9901 & 0.249 & 1.507 & 0.974 & 0.00056 & 1.938\\*[0.5mm]
{\small SoftAdapt ($\boldsymbol{\vartheta}^{\text{\tiny SA}}_1$)} & 0.984 & 0.6187 & 1.3479 & 0.856 & 0.00003614 & 1.436\\*[0.5mm]
\midrule
$\Xi$($\boldsymbol{\vartheta}^{\text{\tiny EW}}_1$) & 0.9\% & 99.86\% & 9.55\% & 38.02\% & 89.025641\% & 13138083\%\\*[0.5mm]
$\Xi$($\boldsymbol{\vartheta}^{\text{\tiny DS}}_1$)& 33.8\% & 285.745\% & 10.181\% & 94.819\% & 37.949\% & 6406079\%\\*[0.5mm]
$\Xi$($\boldsymbol{\vartheta}^{\text{\tiny GN}}_1$) & 0.990\% & 47\% & 9.217\% & 17.349\% & 99.713\% & 12578597\%\\*[0.5mm]
$\Xi$($\boldsymbol{\vartheta}^{\text{\tiny SA}}_1$) & 1.6\% & 31.64\% & 18.801\% & 3.133\% & 99.981\% & 9320338\%\\*[0.5mm]
\bottomrule
\end{tabular}
\vspace*{7.0mm}
\fontsize{10}{12}\selectfont \caption{Reconstructed poroelastic properties in the low-permeability neighborhood $\bxi_2$. Network scaling is applied here.}
\label{gt_vs_reconstruction_NS_LK}
\centering{}%
\vspace*{-2.0mm}
\begin{tabular}{ccccccc}
\toprule
$\boldsymbol{\vartheta}_2$ & $\mu_2$ & $\lambda_2$ & $M_2$ & $\alpha_2$  & $\phi_2$ & $\kappa_2$\\*[0.5mm]
\midrule
{\small ground truth} & 1 & 0.47 & 1.66 & 0.83 & 0.195 & $2.45 \times 10^{-8}$\\*[0.5mm]
\midrule
{\small Equal Weights ($\boldsymbol{\vartheta}^{\text{\tiny EW}}_2$)}\!\!\!\! & 1.001 & 0.4456 & 1.6551 & 0.8395 & 0.0571 & $2.631 \times 10^{-8}$\\*[0.5mm]
{\small DynScl ($\boldsymbol{\vartheta}^{\text{\tiny DS}}_2$)} & 1.0036 & 0.4652 & 1.6596 & 0.8306 & 0.191 & $2.934 \times 10^{-8}$\\*[0.5mm]
{\small GradNorm ($\boldsymbol{\vartheta}^{\text{\tiny GN}}_2$)} & 1.002 & 0.4713 & 1.569 & 0.833 & 0.262 & $2.485 \times 10^{-8}$\\*[0.5mm]
{\small SoftAdapt ($\boldsymbol{\vartheta}^{\text{\tiny SA}}_2$)} & 0.9998 & 0.4702 & 2.3189 & 0.8297 & 0.2164 & $2.0235 \times 10^{-8}$\\*[0.5mm]
\midrule
$\Xi$($\boldsymbol{\vartheta}^{\text{\tiny EW}}_2$) & 0.001\% & 5.2\% & 0.3\% & 1.1\% & 71\% & 7.4\%\\*[0.5mm]
$\Xi$($\boldsymbol{\vartheta}^{\text{\tiny DS}}_2$) & 0.36\% & 1\% & 0.024\% & 0.072\% & 2.1\% & 20\%\\*[0.5mm]
$\Xi$($\boldsymbol{\vartheta}^{\text{\tiny GN}}_2$) & 0.02\% & 0.27\% & 5.5\% & 0.036\% & 35\% & 1.4\%\\*[0.5mm]
$\Xi$($\boldsymbol{\vartheta}^{\text{\tiny SA}}_2$) & 0.02\% & 0.043\% & 40\% & 0.036\% & 11\% & 17\%\\*[0.5mm]
\bottomrule
\end{tabular}
\vspace*{-4.0mm}
\end{table}

In both neighborhoods, dynamic scaling seems to stabilize the training process and improves the network predictions compared to GradNorm and SoftAdapt. This is in particular evident from Fig.~\ref{DW_combined_k2}~(b) where all weighted loss components according to DynScl appear to converge at an approximately similar rates. This is consistent with our analysis at the end of Section~\ref{MP}. SoftAdapt appears to be generally unstable. This may be observed from the loss trajectory in~Figs.~\ref{SA_higher_k} and~\ref{SA_lower_k} and the fact that the network prediction for porosity $\phi^{\text{\tiny SA}}_1\!$ and $\phi^{\text{\tiny SA}}_2\!$ does not converge in both cases when the training process is apparently stalled. GradNorm demonstrates greater stability in terms of the loss behavior. This balancing scheme, however, runs the risk of diminishing the loss sensitivity to a few model parameters. This, for instance, has led to the lack of convergence of porosity $\phi^{\text{\tiny GN}}_1\!$ in Fig.~\ref{GN_lower_k}~(c), and Biot modulus $M^{\text{\tiny GN}}_2\!$ in Fig.~\ref{GN_higher_k}~(e).

\vspace*{-1.0mm}
\subsection{Reconstruction from noisy data}\label{ND}

\begin{figure}[bp!]
\vspace*{-4.0mm}
\begin{centering}
\includegraphics[width=0.85\columnwidth]{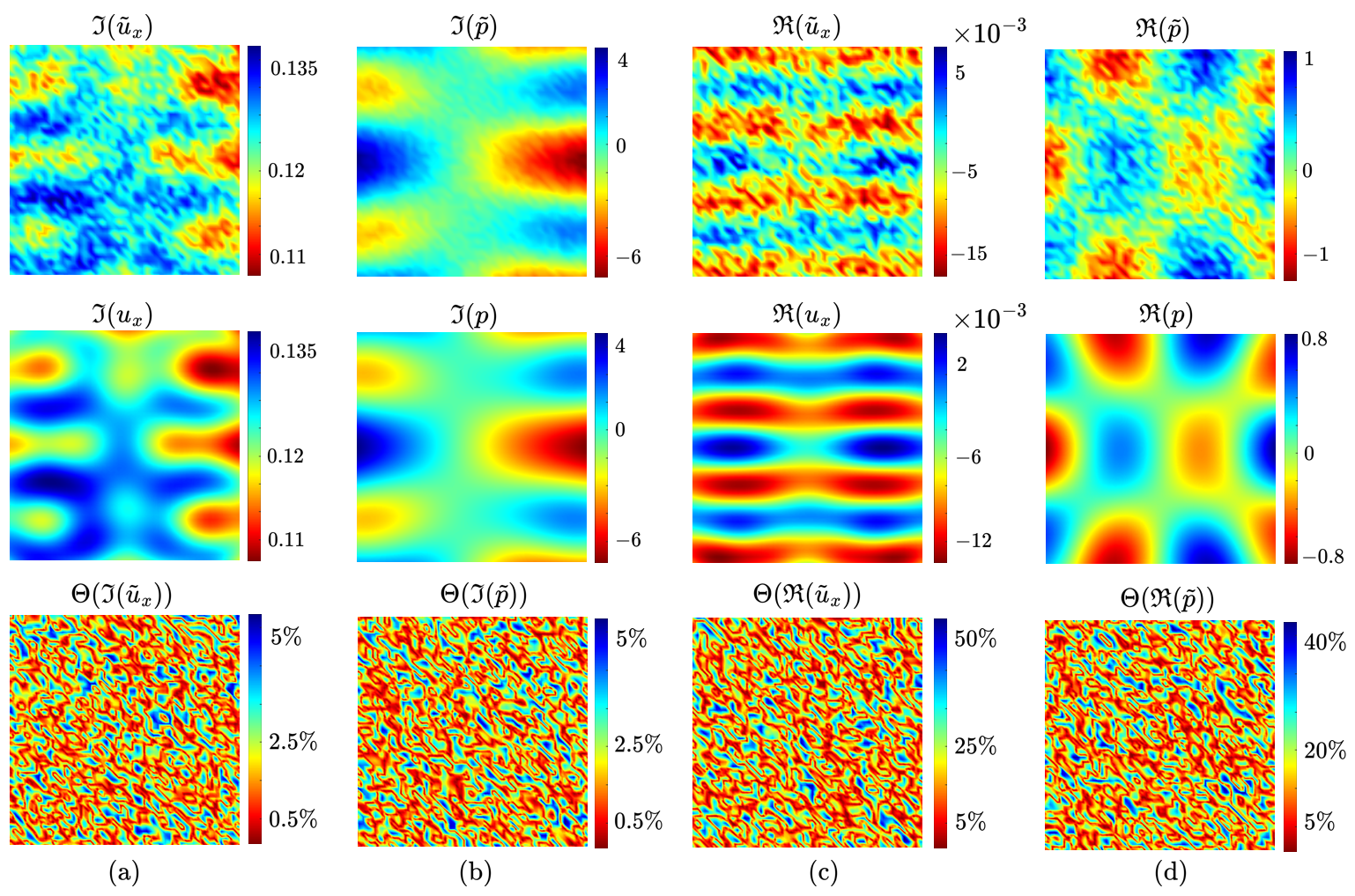}\vspace*{-2.0mm}
\par\end{centering}
\caption{Normal misfit between the noiseless and noisy field components when the overall signal-to-noise ratio is 20:~(a) imaginary part of solid displacement in $x$ direction, (b) imaginary part of pore pressure, (c) real part of solid displacement in $x$ direction, and (d) real part of pore pressure. In each column, top row is the noisy wave field, middle row is its noiseless counterpart, and the bottom row is the normalized different according to~\eqref{normal_misfit_data2}.}
\label{noise_single}
\vspace*{-4.0mm}
\end{figure}

This section aims to investigate the impact of noise in data on the poroelastic properties recovered by way of the proposed approach for model scaling and loss balancing. In this vein, the field data in Section~\ref{NLD} in focal neighborhood $\bxi_1$ is perturbed by $5\%$ noise as the following 
\begin{equation}\label{normal_misfit_data}
\begin{aligned}
&{\tilde{\bZ}} ~=~  \bZ \,+\, \mathfrak{n} \exs \max\left(\left|\bZ\right|\right) \left( I_{\exs\text{N}_x \exs\!\times\exs \text{N}_y} +\, \text{i} \exs I_{\exs\text{N}_x \exs\!\times\exs \text{N}_y} \right),  \quad \mathfrak{n} ~=~0.05,\\*[0.5mm]
&{{\bZ}}(r,s) ~=~ z(x_r,y_s), \quad  z ~=~ u_x,\exs u_y,\exs p, \quad r ~=~ 1,\ldots, \text{N}_x, \quad s ~=~ 1,\ldots, \text{N}_y,
\end{aligned}
\end{equation} 
where $\text{N}_x$ and $\text{N}_y$ are respectively the number of samples in $x$ and $y$ directions in the high-permeability neighborhood $\bxi_1$. Note that introducing the noise as  in~\eqref{normal_misfit_data} as a percentage of the signal range is consistent with how measurement noise is quantified in practice. In multiscale systems, however, these perturbations are not equally reflected on the real and imaginary parts of the field variables. This is shown in Fig.~\ref{noise_single} where the normal misfit between the noiseless and noisy fields are demonstrated for the decomposed solid displacement $u_x$ and pressure $p$ according to  
\begin{equation}\label{normal_misfit_data2}
\begin{aligned}
&\Theta({\tilde{\bZ}}) ~=~\frac{\left| {\tilde{\bZ}} \,-\, \bZ \right|}{\max\left(\left|\bZ\right|\right)}, \quad {{\bZ}}(r,s) ~=~ z(x_r,y_s), \quad z ~=~ \mathfrak{R}(u_x), \mathfrak{I}(u_x), \mathfrak{R}(u_y), \mathfrak{I}(u_y), \mathfrak{R}(p), \mathfrak{I}(p), \\*[0.5mm]
& \hspace{4.5cm} r ~=~ 1,\ldots, \text{N}_x, \quad s ~=~ 1,\ldots, \text{N}_y.
\end{aligned}
\end{equation}  

Observe that the misfit in the major components of $u_x$ and $p$, namely:~$\mathfrak{I}(u_x)$ and $\mathfrak{I}(p)$, reflects the added $5\%$ noise. However, the normal difference for the minor components $\mathfrak{R}(u_x)$ and $\mathfrak{R}(p)$ could surpass $50\%$. This indicates that in multiscale systems, $5\%$ noise in the signal (or data) could imply more than $50\%$ of misfit on certain components of the wave field. It is evident from~\eqref{eq:l1_expanded} and~\eqref{eq:l3} that all components of the field variables and their spatial derivatives actively participate in the loss function~\eqref{eq:lossfunction} and depending on the frequency of wave motion the contribution of highly distorted components by noise could be significant. Thus, even moderate levels of noise could present a critical challenge in multiphysics system identification. This issue is not as problematic in one-scale physical systems. For instance, it is illustrated in our recent study~\cite{xu2023} using both synthetic and experimental data that in elastography of composites i.e., when the PDE system is the (one-scale) Navier equations, $5\%$ noise in data can be easily addressed using the existing tools of signal processing. To address this issue there are several options. For instance, one may switch to the weak (or other energy) form of the governing equations and take advantage of inherent averaging and thus smoothing nature of integral equations, e.g., see~\cite{schm2024,aqui2019,bonn2024,bell2017} for this type of formulation that so far has been implemented in the case of one-scale PDE systems. Here the choice of test functions and boundary contributions will be the main questions. Another approach that is more consistent with the strong formulation of loss function, which is adopted in this study, is to take advantage of signal averaging couple with spectral denoising and FFT-based differentiation of the focal fields. The latter two methods are already deployed in the processing of noiseless signals in Section~\ref{NLD} and a detailed description of them may be found in~\cite{xu2023,pour2018}. For averaging, we assume each test could be repeated $\text{N}_{\text{T}}\!$ times so that the noisy focal fields $\tilde{z}_t$, $t = 0,\ldots, \text{N}_{\text{T}}\!$ can be averaged and used for training in the following form   
\begin{equation}\label{normal_misfit_data3}
\left\langle\exs \tilde{z} \exs\right\rangle_{\text{\tiny N}_{\text{\tiny T}}\!} ~=~\frac{1}{\text{N}_{\text{T}}\!} \sum_{t = 1}^{\text{N}_{\text{T}}\!} \tilde{z}_t, \quad \tilde{z}_t ~=~ \mathfrak{R}(\tilde{u}_{x_t}), \mathfrak{I}(\tilde{u}_{x_t}), \mathfrak{R}(\tilde{u}_{y_t}), \mathfrak{I}(\tilde{u}_{y_t}), \mathfrak{R}(\tilde{p}_t), \mathfrak{I}(\tilde{p}_t),
\end{equation} 

In what follows, we set $\text{N}_{\text{T}}\! =$ 250,$\exs$1500,$\exs$2500. In laboratory experiments, signal averaging is typically set anywhere between 100 to 1000 averages in every experiment~\cite{xu2023,pour2018}. We also selected a significantly larger number of realizations at $\text{N}_{\text{T}}\! =$ 2500 to examine whether by using a sufficiently large ensemble, one could push the reconstruction error arbitrary close to zero regardless of practical considerations.  

\begin{figure}
\begin{centering}
\includegraphics[width=0.95\columnwidth]{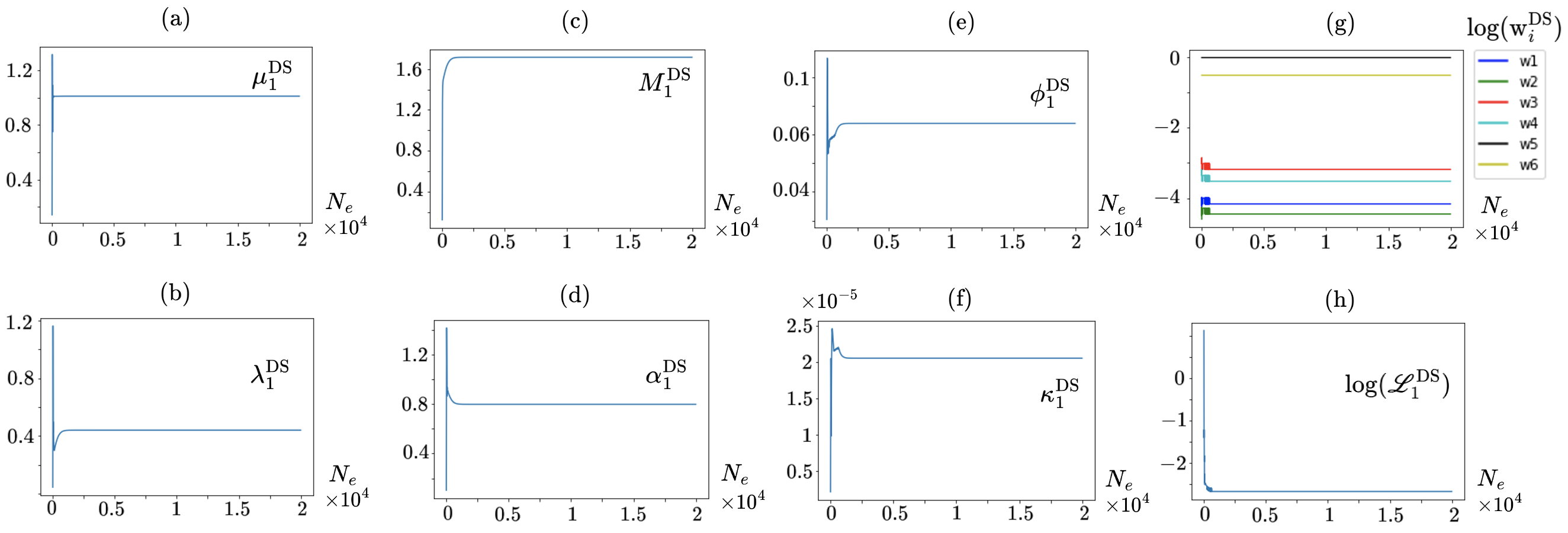}\vspace*{-2.0mm}
\par\end{centering}
\caption{Network-predicted poroelastic properties~\emph{vs.}~the number of epochs $N_{e}$ in the high-permeability neighborhood $\bxi_1$. The network is a scaled MLP and trained by noisy waveforms that are averaged $\text{N}_{\text{T}}\! = 250$ times. The loss is balanced using the dynamic scaling (DS) approach:~(a) drained shear modulus $\mu^{\text{\tiny DS}}_1\!$, (b)~drained first Lam\'{e} parameter $\lambda^{\text{\tiny DS}}_1\!$, (c)~Biot modulus $M^{\text{\tiny DS}}_1\!$, (d)~Biot effective stress coefficient $\alpha^{\text{\tiny DS}}_1\!$, (e)~porosity $\phi^{\text{\tiny DS}}_1\!$, (f)~permeability coefficient $\kappa^{\text{\tiny DS}}_1\!$, (g) DS weights $\text{w}^{\text{\tiny DS}}_i$, $i = 1, \ldots, 6$, versus the minimizer step $N_{e}$, (h) weighted total loss $\text{log}(\mathscr{L}_1^{\text{\tiny DS}})$~\emph{vs.}~epoch.}
\label{noise_250_DW}
\vspace*{-4.0mm}
\end{figure}

\begin{figure}
\begin{centering}
\includegraphics[width=0.95\columnwidth]{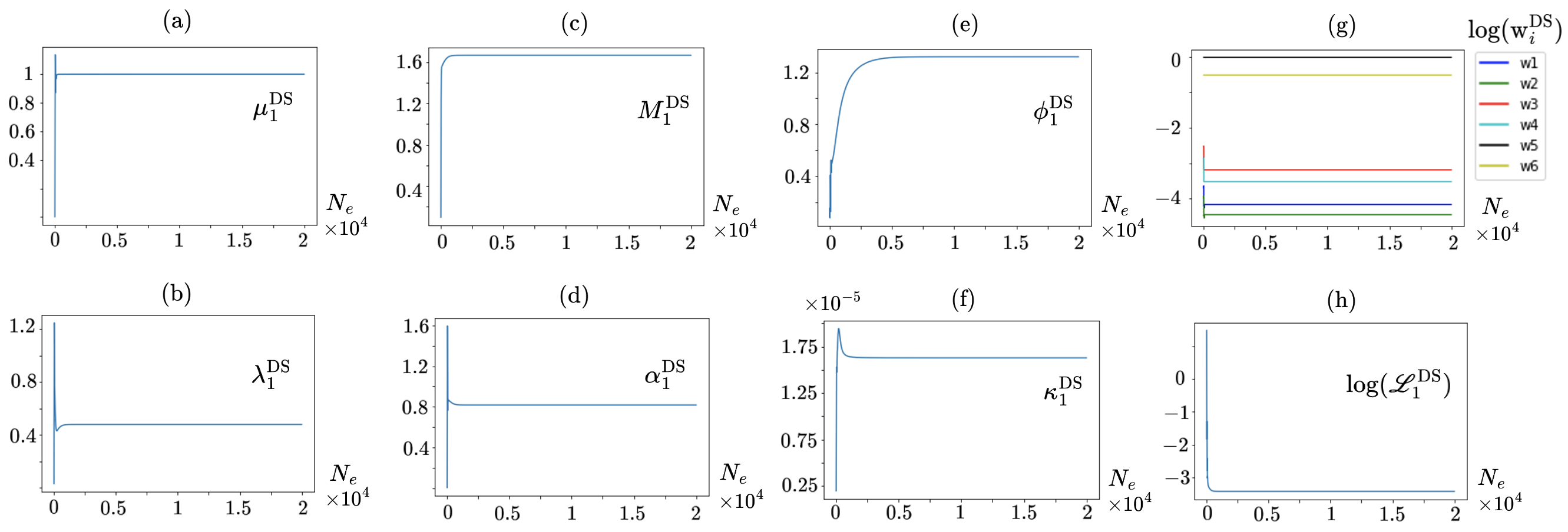}\vspace*{-2.0mm}
\par\end{centering}
\caption{Network-predicted poroelastic properties~\emph{vs.}~the number of epochs $N_{e}$ in the high-permeability neighborhood $\bxi_1$. The network is a scaled MLP and trained by noisy waveforms that are averaged $\text{N}_{\text{T}}\! = 1500$ times. The loss is balanced using the dynamic scaling (DS) approach:~(a) drained shear modulus $\mu^{\text{\tiny DS}}_1\!$, (b)~drained first Lam\'{e} parameter $\lambda^{\text{\tiny DS}}_1\!$, (c)~Biot modulus $M^{\text{\tiny DS}}_1\!$, (d)~Biot effective stress coefficient $\alpha^{\text{\tiny DS}}_1\!$, (e)~porosity $\phi^{\text{\tiny DS}}_1\!$, (f)~permeability coefficient $\kappa^{\text{\tiny DS}}_1\!$, (g) DS weights $\text{w}^{\text{\tiny DS}}_i$, $i = 1, \ldots, 6$, versus the minimizer step $N_{e}$, (h) weighted total loss $\text{log}(\mathscr{L}_1^{\text{\tiny DS}})$~\emph{vs.}~epoch.}
\label{noise_1500_DW}
\vspace*{-4.0mm}
\end{figure}

\begin{figure}
\begin{centering}
\includegraphics[width=0.95\columnwidth]{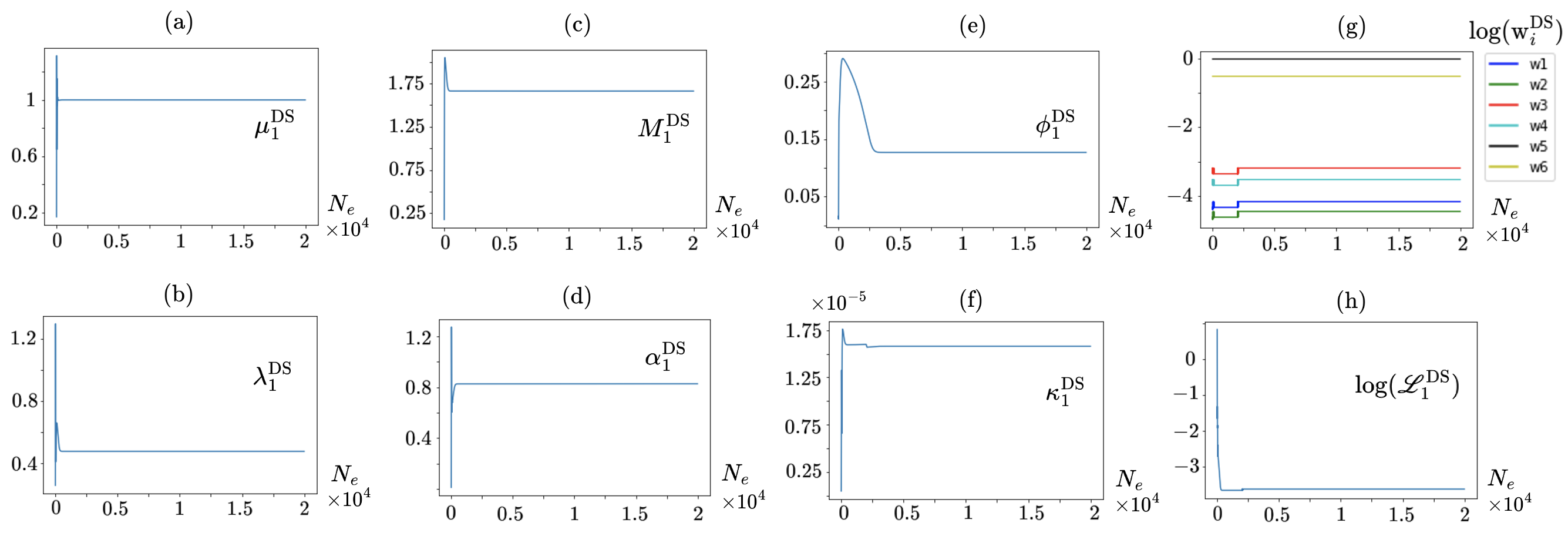}\vspace*{-2.0mm}
\par\end{centering}
\caption{Network-predicted poroelastic properties~\emph{vs.}~the number of epochs $N_{e}$ in the high-permeability neighborhood $\bxi_1$. The network is a scaled MLP and trained by noisy waveforms that are averaged $\text{N}_{\text{T}}\! = 2500$ times. The loss is balanced using the dynamic scaling (DS) approach:~(a) drained shear modulus $\mu^{\text{\tiny DS}}_1\!$, (b)~drained first Lam\'{e} parameter $\lambda^{\text{\tiny DS}}_1\!$, (c)~Biot modulus $M^{\text{\tiny DS}}_1\!$, (d)~Biot effective stress coefficient $\alpha^{\text{\tiny DS}}_1\!$, (e)~porosity $\phi^{\text{\tiny DS}}_1\!$, (f)~permeability coefficient $\kappa^{\text{\tiny DS}}_1\!$, (g) DS weights $\text{w}^{\text{\tiny DS}}_i$, $i = 1, \ldots, 6$, versus the minimizer step $N_{e}$, (h) weighted total loss $\text{log}(\mathscr{L}_1^{\text{\tiny DS}})$~\emph{vs.}~epoch.}
\label{noise_2500_DW}
\vspace*{-4.0mm}
\end{figure}

Table~\ref{gt_vs_reconstruction_HK_noise} presents the reconstruction results from noisy data and associated normal errors, with respect to the ground truth, in the high-permeability neighborhood $\bxi_1$. Here, the property map remains the scaled MLP shown in Fig.~\ref{direct_inversion1}. The proposed dynamic scaling for loss balancing is implemented. The reconstruction results are then accordingly denoted by  
$$\boldsymbol{\vartheta}^{\text{\tiny DS}}_{\nxs\text{\tiny <1>}_{\text{\tiny N}_{\text{\tiny T}}\!}} ~=~ \{\mu^{\text{\tiny DS}}_1,~\lambda^{\text{\tiny DS}}_1,~M^{\text{\tiny DS}}_1,~\alpha^{\text{\tiny DS}}_1,~\phi^{\text{\tiny DS}}_1,~\kappa^{\text{\tiny DS}}_1 \}_{\nxs\text{\tiny <1>}_{\text{\tiny N}_{\text{\tiny T}}\!}}, \quad \text{N}_{\text{T}}\! \in \{250, 1500, 2500\}. $$
In all cases, the network architecture remains the same. $\mu^{\text{\tiny DS}}_1$ remains very close to the ground truth, with minimal deviation across all $\text{N}_{\text{T}}\!$ values. The normal deviation of $\mu^{\text{\tiny DS}}_1$ from the ground truth decreases from 0.99\% at 250 averages to 0.0795\% at 2500 averages, indicating an improved accuracy with an increase in averaging. $\lambda^{\text{\tiny DS}}_1$ shows a slight underestimation compared to its true value. The relative error of $\lambda^{\text{\tiny DS}}_1$ demonstrates a significant reduction from 6.38\% at 250 averages to 1.14\% at 2500 averages, suggesting a substantial enhancement in precision with larger ensembles. $M^{\text{\tiny DS}}_1$ is close to the ground truth, with minor fluctuations. The relative error of $M^{\text{\tiny DS}}_1$ diminishes from 3.27\% at 250 averages to 0.0881\% at 2500 averages, reflecting a similar trend in the reconstructions. $\alpha^{\text{\tiny DS}}_1$ shows minor deviations from the ground truth. The relative error of $\alpha^{\text{\tiny DS}}_1$ reduces significantly from 3.97\% at 250 averages to 0.435\% at 2500 averages. $\kappa^{\text{\tiny DS}}_1$ values are fairly close to the ground truth at $1.5407 \times 10^{-5}$ with small variations. The relative error decreases from 33.7\% at 250 averages to 2.55\% at 2500 averages, showing a marked improvement in accuracy with increased sample size. $\phi^{\text{\tiny DS}}_1$ exhibits considerable variations from the ground truth, especially at lower $\text{N}_{\text{T}}\!$. The relative error drops from 65.2\% at 250 averages to 34.9\% at 2500 averages. However, the true value was never recovered regardless of the ensemble size. Overall, the analysis of Table~\ref{gt_vs_reconstruction_HK_noise} indicates that the accuracy of network estimates improves with the increase in averaging coupled with signal denoising and FFT-based differentiation. However, these measures do not fundamentally address the challenges of reconstructions from noisy data which could be the subject of future studies. The convergence plots along with network predictions as a function of optimizer step $N_e$ are shown in Figs~\ref{noise_250_DW},~\ref{noise_1500_DW},~\ref{noise_2500_DW} for the reconstructions reported in Table~\ref{gt_vs_reconstruction_HK_noise}. Similar to reconstructions from noiseless data, training by way of dynamic scaling shows stable trajectories for both loss and network predictions.

\begin{table*}
\fontsize{10}{12}\selectfont \caption{Reconstruction of poroelastic properties in the high-permeability neighborhood $\bxi_1$ from noisy data. The waveforms are averaged $\text{N}_{\text{T}}\! \in \{250, 1500, 2500\}$ times. Network scaling is applied here.}
\label{gt_vs_reconstruction_HK_noise}
\centering{}%
\vspace*{-2.0mm}
\begin{tabular}{ccccccc}
\toprule
$\boldsymbol{\vartheta}_1$ & $\mu_1$ & $\lambda_1$ & $M_1$ & $\alpha_1$  & $\phi_1$ & $\kappa_1$\\*[0.75mm]
\midrule
{\small ground truth} & 1 & 0.47 & 1.66 & 0.83 & 0.195 & $1.5407 \times 10^{-5}$\\*[0.75mm]
\midrule
{\small DynScl ($\boldsymbol{\vartheta}^{\text{\tiny DS}}_{\nxs\text{\tiny <1>}_{\text{\tiny 250}}}$)} & 1.00991 & 0.44 & 1.714 & 0.7971 & 0.06786 & $2.06 \times 10^{-5}$\\*[0.75mm]
{\small DynScl ($\boldsymbol{\vartheta}^{\text{\tiny DS}}_{\nxs\text{\tiny <1>}_{\text{\tiny 1500}}}$)}  & 1.00003 & 0.479 & 1.6696 & 0.8175 & 0.13199 & $1.6295 \times 10^{-5}$\\*[0.75mm]
{\small DynScl ($\boldsymbol{\vartheta}^{\text{\tiny DS}}_{\nxs\text{\tiny <1>}_{\text{\tiny 2500}}}$)}  & 0.9992 & 0.475 & 1.6614 & 0.8264 & 0.12678 & $1.5799 \times 10^{-5}$\\*[0.75mm]
\midrule
{\small$\Xi$($\boldsymbol{\vartheta}^{\text{\tiny DS}}_{\nxs\text{\tiny <1>}_{\text{\tiny 250}}}$)} & 0.99\% & 6.38\% & 3.27\% & 3.97\% & 65.2\% & 33.7\%\\*[0.75mm]
{\small$\Xi$($\boldsymbol{\vartheta}^{\text{\tiny DS}}_{\nxs\text{\tiny <1>}_{\text{\tiny 1500}}}$)} & 0.00312\% & 1.98\% & 0.577\% & 1.51\% & 32.3\% & 5.76\%\\*[0.75mm]
{\small$\Xi$($\boldsymbol{\vartheta}^{\text{\tiny DS}}_{\nxs\text{\tiny <1>}_{\text{\tiny 2500}}}$)} & 0.0795\% & 1.14\% & 0.0881\% & 0.435\% & 34.9\% & 2.55\%\\*[0.75mm]
\bottomrule
\end{tabular}
\vspace*{-4.0mm}
\end{table*}

\subsection{Discussion}

Section~\ref{ER} showcases the capability of scaled neural networks in simultaneous reconstruction of poroelastic properties in distinct neighborhoods whose permeability coefficients differ by multiple orders of magnitude. This is thanks to the last layer in the proposed architecture which allows the network predictions to straddle a range of potential scales in heterogenous environments and does not require an exact a priori knowledge of the scaling of unknown physical properties. The latter would be required if the scaling were to be hardcoded. Note that in this model, the neural map is factorized into two operators; the first operator, which involves the unknown weights and biases, acts as a unit (or normal) shape function describing the spatial distribution of each property, while the second operator magnifies the first operator's output according to a pre-determined set of likely scales for each physical property. It should be mentioned that in this formulation since the scaling of output is isolated and transferred as a whole to the last layer, the unknown network parameters (weights and biases) can be clamped at O(1) during the optimization process. This accelerate the training and paves the way toward a \emph{physics-based loss balancing approach} through establishing explicit estimates for the scale of each loss component and its gradients with respect to the model parameters. In the case of Biot equations, used for poroelastography, we took advantage of this architecture to show that each weighted loss component is a Lipchitz function of the network weights and the proposed dynamic scaling (DynScl) of the loss -- based on the average scale of each loss component -- ensures that the Lipchitz constant remains of O(1) for all network parameters during training. This manifests itself in three advantages that we observed in the reconstructions from noiseless data, especially when the results were compared with that of Softmax adaptive weights (SoftAdapt) and gradient normalization (GradNorm). The first advantage is \emph{stability and robustness}; compared to both SoftAdapt and GradNorm, the training by DynScl weights is more stable in that (a) all weighted loss components and the total loss converge with a unified rate, (b) the proper weight for each objective is quickly identified and remains stable throughout the optimization process, (c) the total loss converges to zero, and (d) all the network-predicted properties show convergence to certain limits. (a) and (b) remain the case in GradNorm, even though the estimated loss weights by the DynScl and GradNorm differ by orders of magnitude. Keep in mind that GradNorm optimizes the weights in each epoch in order to unify the rate of change of weighted loss components with respect to a shared layer in the neural network (not all the network parameters due to computational considerations). In Section~\ref{ER}, all GradNorm-estimated weights are of O(1), in both focal regions, which could explain the remarkably large magnitude of the total loss and its potential insensitivity to the physics at smaller scales. In this case, while the total loss uniformly decreases, we observe that a few network-estimated parameters related to the physics of fluid flow in the pore space do not converge i.e.,~(c) and (d) do not seem to hold in GradNorm-based balancing. Another point worth mentioning is the impact of noise in data on the loss balancing procedure. Note that DynScl only requires an estimated scale for the field derivatives. While direct FFT-based differentiation of focal fields provides the best scaling estimate, as implemented in this study, this is not generally required and an estimate on the scale of derivatives can be obtained by an elementary FFT analysis of the measurements coupled with network-predicted scales for the physical properties at every epoch. GradNorm relies on the estimated derivatives furnishes by automatic differentiation to optimize the loss weights at every epoch. The latter when combined with noisy or incomplete data may lead to instability as reported by~\cite{amin2023}. In this study, since the data is dense and the waveforms were carefully processed prior to training for any of the loss balancing techniques, this problem was not observed. SoftAdapt is generally unstable in the reconstructions of Section~\ref{ER}. We believe that this is due to the limited range and sensitivity of the Softmax function for multiphysics objectives. Varying between zero and one, SoftAdapt weights are assigned such that objectives with small decay rates are more visible to the optimizer. In multiphysics systems, the decay rates are largely influenced by the scale of each loss component such that a fast converging task could assume a much smaller decay rate compared to a slow converging objective of larger magnitude. This could confuse or mislead the optimizer to focus on the objectives that are already converging and reduce the loss sensitivity to certain outputs. This may explain the behavior observed in the SoftAdapt convergence plots wherein while the loss converges, some of the network-predicted properties are diverging. The second advantage of the proposed approach is in its \emph{flexibility and efficiency}. Network scaling is a quite simple idea and can be applied to any architecture. Moreover, DynScl weights can be explicitly computed for both Lipschitz and non-Lipschitz objectives (see Remark~\ref{rm1}) and does not require a parallel optimization process per epoch. In this respect, the computational cost of DynScl and SoftAdapt is similar. The third advantage of loss balancing by way of dynamic scaling is \emph{clarity}. The DynScl weights are directly driven by the physics of each objective such that each loss component represents the balance of physical quantities of $O(1)$. In this framework, the total loss converges to zero ensuing that all objectives are met. In reconstructions of Section~\ref{ER}, the SoftAdapt and GradNorm weights are all of O(1) and do not seem to be cognizant of the physics of loss components. This resulted in quite large-scale total losses that while significantly decrease by multiple orders during training, they do not converge to zero (or a near-zero value), causing an ambiguity in interpretation of the results as to whether the global convergence is achieved and all objectives are sufficiently met. In the reconstructions from noisy data, we showed that DynScl weights remain stable. We also implemented signal averaging coupled with spectral denoising and FFT-based differentiation to improve the reconstructions. The results suggest that while the proposed measures significantly enhance the accuracy of network predictions, they do not fully address the challenges involved in multiphysics system identification from noisy data. These challenges seem to be more related to errors in the calculation of derivatives and the question of observability of small-scale physics in presence of noise, rather than loss balancing which could be the subject of a future work.            

\section{Conclusion}

An intelligent framework is established for identification of multiphysics systems from (reconstructed or measured) local wave fields. As an application, we focused on comprehensive characterization of poroelastic materials from solid displacement and pore pressure data. To this end, the neural maps, predicting the six unknown parameters of Biot equations, are constructed by a normal map describing the spatial distribution of each property that is composed into a scaling layer. This architecture allows its various outputs to assume a set of likely scales and does not require an exact knowledge of the scaling of unknown physical properties. In this model the unknown network parameters in the normal map remain of O(1) during training. This forms the basis for physics-based balancing of the loss function through the proposed dynamic scaling. The latter is formulated separately for Lipschitz and non-Lipschitz objectives (with respect to the network's weights and biases) through establishing explicit estimates for the scale of each loss component and its gradients with respect to the model parameters. The proposed approach that integrates the idea of network scaling with scale-driven loss balancing is then put to test by a set of numerical experiments. Special attention is paid to tight formations (low-permeability regions) and simultaneous reconstruction of all PDE parameters in highly heterogeneous environments were some unknown properties could straddle multiple scales in various neighborhoods. We demonstrated that the proposed method is successful in addressing some of these challenges. A comparative analysis is conducted with two state-of-the-art techniques for loss balancing, namely: GradNorm and SoftAdapt. We demonstrated both logically and computationally that scale-based modeling and loss balancing offer three advantages in terms of stability, efficiency, and accuracy that may be particularly relevant in multiphysics data processing. The proposed method is formulated in a generic platform so that its application to other physical systems is straightforward.

\section*{Acknowledgments} 

This study was funded by the National Science Foundation (Grant No.~1944812) and the University of Colorado Boulder through Fatemeh Pourahmadian's startup. This work deployed resources from the University of Colorado Boulder Research Computing Group, which is supported by the National Science Foundation (awards ACI-1532235 and ACI-1532236), the University of Colorado Boulder, and Colorado State University. Special thanks are due to Kevish Napal for facilitating the use of FreeFem++ code developed as part of~\citep{pour2022} for poroelastodynamic simulations.

\appendix

\section{Loss components}\label{apx} 
The expanded loss components $\ell_2$ to $\ell_6$ in~\eqref{eq:l1} are listed in the following.
\begin{equation}
\label{eq:l3}
\begin{split}
 \ell_2 ~=~ \mu&\left(\frac{\partial^2 \mathfrak{I}(u_x)}{\partial x^2}\,+\,\frac{\partial^2 \mathfrak{I}(u_x)}{\partial y^2}\right)\,+\,(\lambda+\mu)\left(\frac{\partial^2 \mathfrak{I}(u_x)}{\partial x^2}\,+\,\frac{\partial^2 \mathfrak{I}(u_y)}{\partial x \partial y}\right) \,-\, \\
& \,\,\,\, -\mathfrak{R}(a) \frac{\partial \mathfrak{I}(p)}{\partial x}\,-\,\mathfrak{I}(a) \frac{\partial \mathfrak{R}(p)}{\partial x}\,+\,\omega^2 \mathfrak{R}(b) \mathfrak{I}(u_x)\,+\,\omega^2 \mathfrak{I}(b) \mathfrak{R}(u_x) \,-\, \mathfrak{I}(f^{u}_x),\\*[2 mm]
 \ell_3 ~=~  \mu&\left(\frac{\partial^2 \mathfrak{R}(u_y)}{\partial x^2}\,+\,\frac{\partial^2 \mathfrak{R}(u_y)}{\partial y^2}\right)\,+\,(\lambda+\mu)\left(\frac{\partial^2 \mathfrak{R}(u_y)}{\partial y^2} \,+\, \frac{\partial^2 \mathfrak{R}(u_x)}{\partial x \partial y}\right) \,-\, \\
& \,\,\,\, -\mathfrak{R}(a) \frac{\partial \mathfrak{R}(p)}{\partial y} \,+\,\mathfrak{I}(a) \frac{\partial \mathfrak{I}(p)}{\partial y}\,+\,\omega^2 \mathfrak{R}(b) \mathfrak{R}(u_y)\,-\,\omega^2 \mathfrak{I}(b) \mathfrak{I}(u_y)  \,-\, \mathfrak{R}(f^{u}_y),\\*[2 mm]
 \ell_4 ~=~  \mu&\left(\frac{\partial^2 \mathfrak{I}(u_y)}{\partial x^2}\,+\,\frac{\partial^2 \mathfrak{I}(u_y)}{\partial y^2}\right)\,+\,(\lambda+\mu)\left(\frac{\partial^2 \mathfrak{I}(u_y)}{\partial y^2} \,+\, \frac{\partial^2 \mathfrak{I}(u_x)}{\partial x \partial y}\right)  \,-\, \\
& \,\,\,\, -\mathfrak{R}(a) \frac{\partial \mathfrak{I}(p)}{\partial y} \,-\, \mathfrak{I}(a) \frac{\partial \mathfrak{R}(p)}{\partial y}\,+\,\omega^2 \mathfrak{R}(b) \mathfrak{I}(u_y)\,+\,\omega^2 \mathfrak{I}(b) \mathfrak{R}(u_y)  \,-\, \mathfrak{I}(f^{u}_y),\\*[2 mm]
&\hspace{-11mm} \ell_5 ~=~  \mathfrak{R}(a)\left(\frac{\partial \mathfrak{R}(u_x)}{\partial x}\,+\,\frac{\partial \mathfrak{R}(u_y)}{\partial y}\right)\,-\,\mathfrak{I}(a)\left(\frac{\partial \mathfrak{I}(u_x)}{\partial x}\,+\,\frac{\partial \mathfrak{I}(u_y)}{\partial y}\right) \,+\, \\
&\hspace{7mm} \,+\, \frac{\mathfrak{R}(c)}{\omega^2} \left(\frac{\partial {\mathfrak{R}(p)}}{\partial x^2}\,+\,\frac{\partial {\mathfrak{R}(p)}}{\partial y^2}\right)\,-\,\frac{\mathfrak{I}(c)}{\omega^2}  \left(\frac{\partial \mathfrak{I}(p)}{\partial x^2}\,+\,\frac{\partial \mathfrak{I}(p)}{\partial y^2}\right)\,+\,\frac{\mathfrak{R}(p)}{M} \,+\, \mathfrak{R}(c) \frac{\mathfrak{R}(f^p)}{\omega^{2}} \,-\, \mathfrak{I}(c) \frac{\mathfrak{I}(f^{p})}{\omega^{2}}, \\*[2.5 mm]
&\hspace{-11mm} \ell_6 ~=~  \mathfrak{R}(a)\left(\frac{\partial \mathfrak{I}(u_x)}{\partial x}\,+\,\frac{\partial \mathfrak{I}(u_y)}{\partial y}\right)\,+\,\mathfrak{I}(a)\left(\frac{\partial \mathfrak{R}(u_x)}{\partial x}\,+\,\frac{\partial \mathfrak{R}(u_y)}{\partial y}\right) \,+\,  \\
&\hspace{7mm}  \,+\, \frac{\mathfrak{R}(c)}{\omega^2} \left(\frac{\partial \mathfrak{I}(p)}{\partial x^2} \,+\, \frac{\partial \mathfrak{I}(p)}{\partial y^2}\right) \,+\,\frac{\mathfrak{I}(c)}{\omega^2} \left(\frac{\partial \mathfrak{R}(p)}{\partial x^2}\,+\,\frac{\partial \mathfrak{R}(p)}{\partial y^2}\right) \,+\, \frac{\mathfrak{I}(p)}{M} \,+\, \mathfrak{R}(c) \frac{\mathfrak{I}(f^p)}{\omega^{2}} \,+\, \mathfrak{I}(c) \frac{\mathfrak{R}(f^{p})}{\omega^{2}},
\end{split}
\end{equation}

\bibliography{paper2_v3}

\end{document}

%% file: macros.tex
\newlength{\kaka}

\newcommand{\ahref}[2]{}

\newcommand{\beq}{\begin{equation}}
\newcommand{\eeq}{\end{equation}}
\newcommand{\lb}{\label}

\newcommand{\bea}{\begin{eqnarray}}
\newcommand{\eea}{\end{eqnarray}}
\newcommand{\bxr}{\begin{array}}
\newcommand{\exr}{\end{array}}

\newcommand\exs{\hspace*{0.4mm}}

\newcommand\nxs{\hspace*{-0.2mm}}

\newcommand{\norms}[1]{\parallel\! #1 \!\parallel}

\newcommand{\bZ} {\boldsymbol{Z}}

\newcommand{\bx} {\boldsymbol{x}}

\newcommand{\bzero}{\boldsymbol{0}}

\newcommand{\bu} {\boldsymbol{u}}

\newcommand{\bxi} {\boldsymbol{\xi}}